\documentclass[preprint,10pt]{elsarticle}             
\journal{Combustion and Flame}

\usepackage[latin1]{inputenc}
\usepackage{graphicx}
\graphicspath{{./Figs/}}
\usepackage{ulem}
\usepackage{comment}
\usepackage{grffile}	 
\usepackage{tabularx}
\usepackage{booktabs}
\usepackage{caption}
\usepackage[mathscr]{euscript}

\usepackage{amsmath}
\usepackage{amssymb}
\usepackage{mathtools}

\usepackage{geometry}
\usepackage{lineno}

\usepackage[version=4]{mhchem}

\usepackage{xspace}

\newcommand{\lra}{\ensuremath{\leftrightarrow}\xspace}
\newcommand{\C}[1]{\ensuremath{[\ce{#1}]}}
\newcommand{\nox}{\ensuremath{\ce{NO_x}}\xspace}
\newcommand{\radicals}{\ce{O}, \ce{OH} and \ce{H}\xspace}

\usepackage[nodots]{numcompress}

\usepackage{lineno}

\biboptions{square, numbers, sort&compress}
\usepackage{color}
\usepackage[pdftex,bookmarks,colorlinks,breaklinks]{hyperref}
\definecolor{dullmagenta}{rgb}{0.4,0,0.4}   
\definecolor{darkblue}{rgb}{0,0,0.4}
\definecolor{darkred}{rgb}{.9,0.,0.}
\definecolor{darkgreen}{rgb}{0.1,0.7,0.0}   
\hypersetup{linkcolor=red,citecolor=blue,filecolor=dullmagenta,urlcolor=darkblue}

\newlength{\figurewidth}
\begin{document}


    \begin{frontmatter}
    
    \author[Kentucky]{Jos\'e Gra\~na-Otero\corref{Contact}}
    \ead{jose.c.grana@gmail.com}
    
    \address[Kentucky]{Dept. of Mechanical Engineering, University of Kentucky. \\ 279 RGAN Building, Lexington, KY. USA. Phone: (859) 218-0645}
    \cortext[Contact]{Corresponding author}

    \title{The recombination region in lean steady premixed \ce{H2} flames.}



%
%
%
%
%


\begin{abstract}

\ce{H2} premixed flames are well known for a long, trailing region where the unburnt \ce{H2} and the super-equilibrium concentrations of radicals left past the fuel consumption layer gradually decay to thermodynamic equilibrium. This recombination region, it's argued here, is a second order effect induced by the premature quenching of the shuffle reactions, which inhibits the decay to equilibrium in the first approximation. 



Its structure and kinetics are studied in detail to capture the small but finite reaction rates accounting for its characteristic length scale, which is large enough to render the diffusive transport negligible, hence deactivating the upstream feedback link with the main flame structure. It is isothermal and can be described, in moderately lean flames, by just the distribution of \ce{H2} as the sole degree of freedom, a drastic reduction consequence of the strongly constrained evolution imposed by the almost exactly quenched shuffle reactions. 


The rate of decay to equilibrium of \ce{H2} and radicals is thus dictated by the balance between the convective transport and chemical sinks controlled by the \ce{HO2} kinetics, which becomes dominant after the shuffle reactions quench. Essentially, it is a two-step mechanism with the first, rate-limiting step converting \ce{O2} into \ce{HO2}, whereas the second one, comprising multiple fast, parallel pathways, depletes radicals as \ce{HO2} is converted into \ce{O2} and \ce{H2O}.

Generalizations to very lean and nearly stoichiometric flames, when the one degree of freedom model is not applicable, as well as the effect of heat losses are also briefly discussed.

\end{abstract}

\begin{keyword}
	Hydrogen premixed flames \sep Recombination region \sep Reduced hydrogen kinetics \sep Shuffle reactions \sep Hydroperoxyl kinetics.
\end{keyword}

\end{frontmatter}


\fontsize{10pt}{12pt}\selectfont

\begin{center}
\fbox{
\begin{minipage}{0.96\textwidth}

\quad {\textbf{Nomenclature}}

\begin{tabbing}

\quad $\dot{\mathcal{C}}$  \hspace{0.7cm} \= total molar consumption rate ($kmole/(m^3  s)$). \\

\quad $\mathcal{D}_{\ce{H2}}$  \> $\ce{H2}$ mixture mass-averaged diffusivity ($m^2/s$). \\

\quad $\mathcal{D}$  \> operator $X_{\ce{H2}}^{-1} \partial () /\partial \xi$ defined in \eqref{MainEquationsDimensionless}. \\
	
\quad $\mathcal{E}$ \hspace{0.7cm} \= relative deviation from exact partial equilibrium. \\

\quad $[i]$ \> molar concentration of species $i$ ($kmole/m^3$). \\

\quad $k_{i}$ \> Arrhenius reaction rate constant in units $kmole$, $m^3$ and $s$. \\

\quad $\mathcal{K}_{i}$ \> equilibrium constant of the reaction $i$.. \\

\quad $\mathcal{L}$ \> local length scale of the \ce{H2} distribution. \\

\quad $\mathcal{L}^o$ \> reference length scale \eqref{lengthScale}. \\

\quad $\dot{m}''_L$ \> laminar flame mass consumption flux ($kg/(m^2s)$). \\

\quad $\mathcal{M}_i$ \> molecular weight of species $i$. \\

\quad $\dot{\mathcal{N}}$ \> net molar production rate $\dot{\mathcal{P}} - \dot{\mathcal{C}}$. \\

\quad $\dot{\mathcal{P}}$ \> total molar production rate ($kmole/(m^3 s)$). \\

\quad $Pe_r$ \> Péclet number in the recombination region. \\

\quad $\dot{\mathscr{S}}^e$ \> molar production rate from all out-of-equilibrium reactions ($kmole/(m^3 s)$). \\

\quad $u$ \> flow velocity relative to the flame ($m/s$). \\
 
\quad $x$ \> coordinate along the flame ($mm$). \\

\quad $X$ \> mole fraction. \\

\quad $Y$ \> mass fraction. \\

\quad $\acute{X}$ \> mole fraction derivative with respect to $X_{\ce{H2}}$. \\

\quad $z$ \> reduced \ce{H2} concentration $\C{\ce{H2}}^b/\C{\ce{H2}}$. \\ [.2cm]

\small{Greek:} \\

\quad $\alpha$ \> branching ratio $\dot{\omega}_{17} /\dot{\omega}_{13}$. \\

\quad $\delta_L$ \> flame thickness. \\

\quad $\gamma_{m}$ \> scaled rate $\dot{\omega}'^{e}_{m, f}/\dot{\omega}'^{e}_{13, r}$ of the \ce{HO2} reaction $m$ (\ref{HO2RatesApp}). \\

\quad $\tilde{\Gamma}_4$ \> regular part $\tilde{\omega}_{4}/(1-z^2)$ of the rate of reaction 4 \eqref{Gamma4Source}. \\

%
\quad $\phi$  \> equivalence ratio. \\

\quad $\psi_i$  \> scaled mass fraction $Y_{i}/Y^e_{i}$. \\

%
%
%

\quad $\tilde{\Omega}_i$  \>  scaled reaction rates of the shuffle reactions $\dot{\omega}^{e}_{i,f}/\dot{\omega}^{e}_{13,f}$. \\

\quad $\dot{\omega}_i$  \>  molar reaction rate of reaction $i$ ($kg/m^3$). \\

\quad $\rho$ \> density ($kg/m^3$). \\

\quad $\tau$ \> characteristic time ($s$). \\

\quad $\xi$ \> coordinate $x$ scaled with $\mathcal{L}^o$. \\ [.2cm]

\small{subscripts:} \\

\quad f, r \> forward, reverse. \\ [.2cm]

\small{superscripts:} \\

\quad b \> burnt thermodynamic equilibrium. \\

\quad e \>  shuffle reactions partial equilibrium concentrations \eqref{PartialEquilibriumSolvedOnlyH2}. \\

\quad u \> fresh mixture. \\

\quad $0$ \> beginning of the recombination region. \\ [.2cm]

\small{accents:} \\

\quad $\tilde{ }$ \> rates scaled with $\dot{\omega}_{13f}$. \\

\quad $\dot{ }$ \> magnitude per unit time \hspace{10cm}.

\end{tabbing}
\end{minipage}}
\end{center}

\section{Introduction.}

The structure of flames has often been investigated using an overall irreversible reaction between the fuel and the oxidizer, with a rate proportional to the mass fraction of the limiting reactant and an Arrhenius dependence on the temperature (see for instance the classic references by Ya. B. Zel'dovich \cite{ZeldovichBook}, F. A. Williams \cite{WilliamsBook}, A. Li\~n\'an \cite{Linhan74} and Sivashinsky \cite{SivashiskyHydrodynamics}). The chemical reaction is in this case essentially confined to the so-called reaction or fuel consumption layer. Outside it, the chemical reaction is either frozen, with too small a rate to be significant when the temperature is too low; or in equilibrium, with zero net rate because one of the reactants has been depleted.

In these models, chemical equilibrium is quickly reached right downstream of the reaction layer, once the deficient reactant has been exhausted. Without source terms, the temperature and species remain constant downstream of steady flames, so zero-gradients are the appropriate downstream boundary conditions.

Flames with more complex and realistic kinetic mechanisms can be as well understood using the same basic framework \cite{SeshadriReview}; namely a thin reaction layer, where the chemical activity is confined, flanked by transport regions where reactions are either frozen or in equilibrium. 

However the presence of many competing chemical reactions introduce additional structure inside the fuel consumption layer, with nested and overlapping sub-layers, that complicate extraordinarily the analysis, even in relatively simple cases such as hydrogen \cite{WilliamsHydrogenSanchez, DaniHydrogen, DaniHydrogenI} or methane \cite{seshadriMethane, SeshadriMethaneII, LinanLeanMethane}.

These difficulties are due to the presence of highly reactive intermediate radicals such as \radicals and many others present in complex fuels. Although present in much smaller amounts than the main species, radicals are nevertheless produced in the reaction layer in quantities largely exceeding the values corresponding to equilibrium at the local temperature. 

Most of these intermediate species are typically short-lived and their concentrations decay rapidly outside the fuel consumption layer. 

However, intriguingly, despite their high reactivity in both \ce{H2} and hydrocarbon fuels, \radicals typically decay at much slower rates than most of other radicals and can be found at relatively large distances downstream of the fuel consumption layer in large amounts compared with the final equilibrium values. 

All the same, chemical equilibrium, or equivalently, zero-gradients of temperature and species are still used as boundary conditions right downstream of the fuel consumption layer, somehow ignoring the tail of slow decay of these radicals.


To understand these aspects, this paper analyzes the structure of the high-temperature region -- hereafter referred to as the recombination region --, where \ce{H2} and radicals leaking downstream of the fuel consumption layer decay to the burnt thermodynamic equilibrium state.


%
\begin{figure}[h!]
	\begin{center}
		\includegraphics[width=0.44\textwidth]{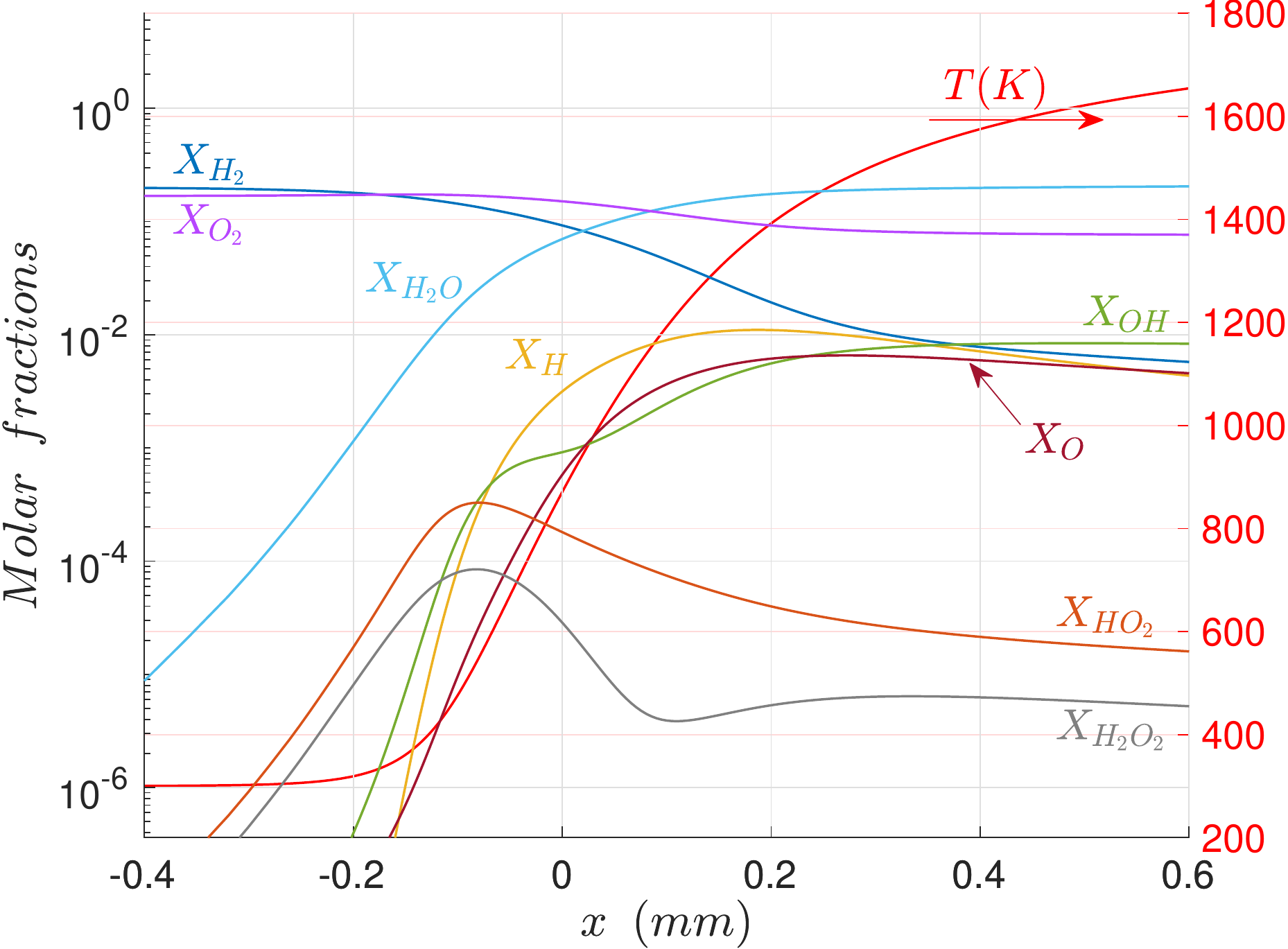}  \hspace{1cm}
		\includegraphics[width=0.44\textwidth]{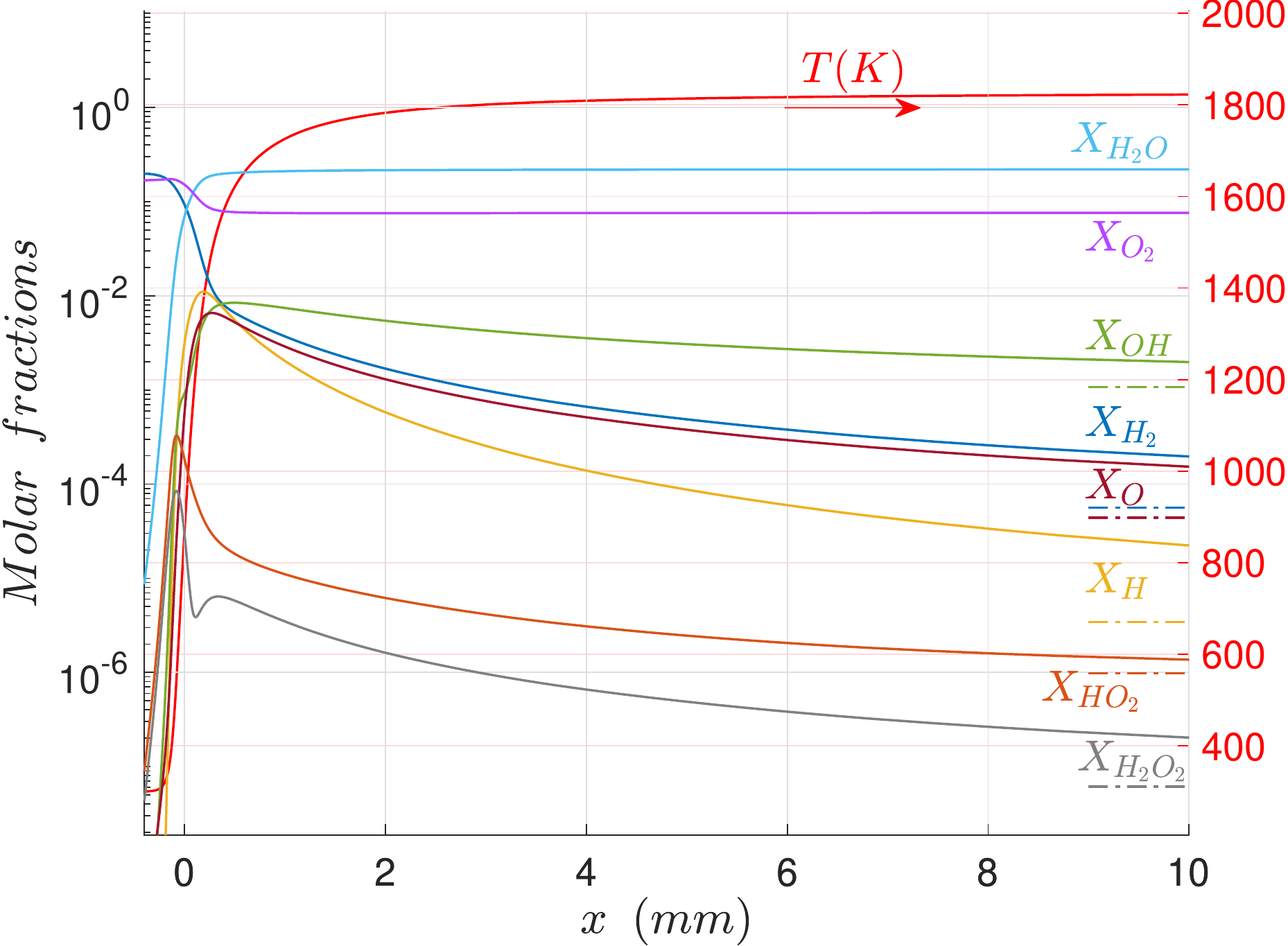}  \\[.2cm]
		\includegraphics[width=0.44\textwidth]{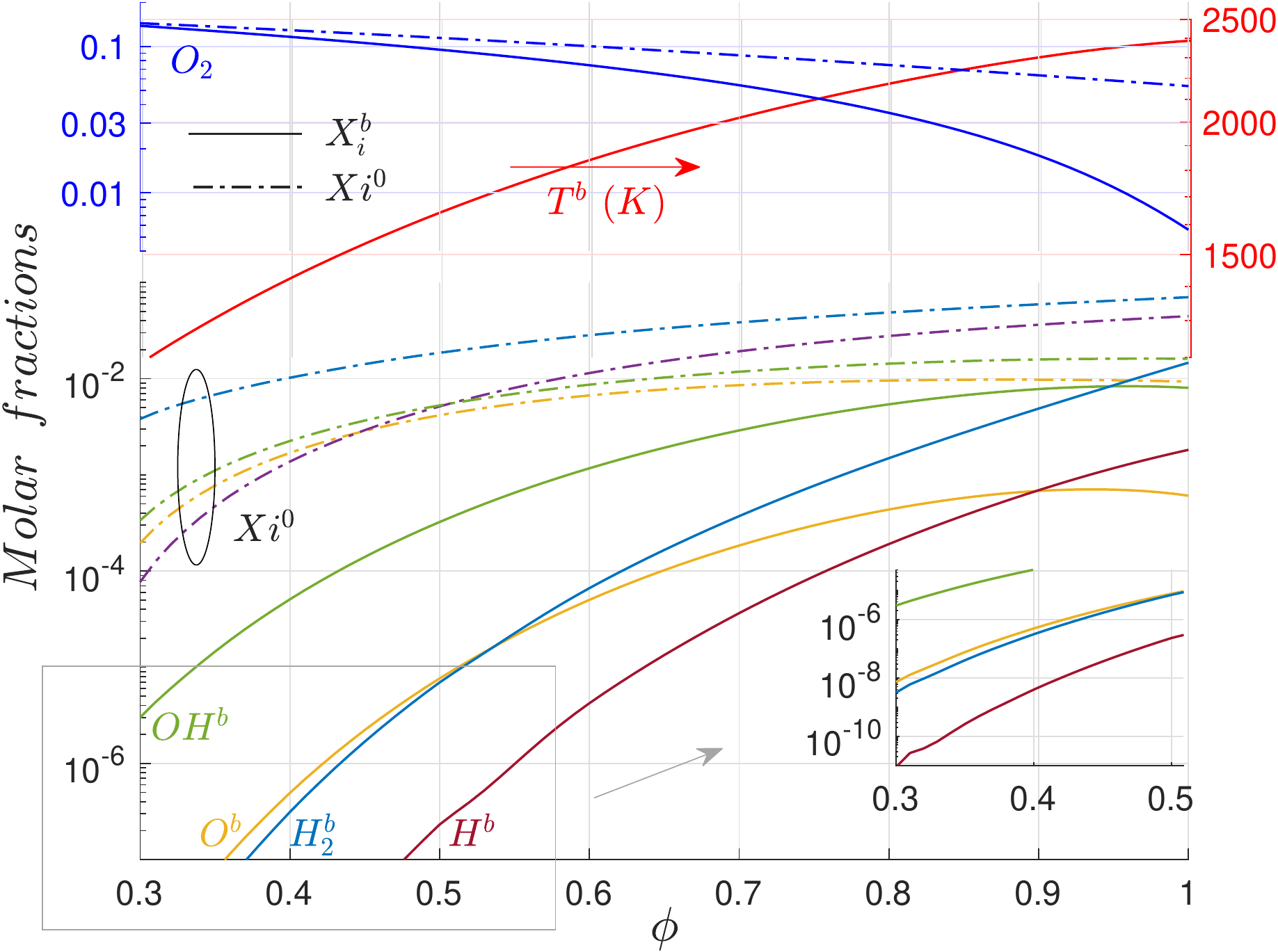}
	\end{center}
	\caption{Distributions, in a $\phi = 0.6$ \ce{H2}-air flame, of the molar fractions $X$ of the main species and radicals in the fuel consumption layer (top left) and in the recombination region (top right) where the burnt thermodynamic equilibrium is finally attained. The values corresponding to this state -- denoted in \texttt{?}he text with the $b$ superscript -- are represented by the small dash-dot segments (indistinguishable from the solid lines in the case of \ce{O2} and \ce{H2O}). The origin of $x$ coincides with the \ce{HO2} peak production rate, found in the fuel consumption layer for all values of the equivalence ratio $\phi$. The bottom plot shows the variation with $\phi$ of the final burnt equilibrium values $b$. The dash-dot lines represent the corresponding values $X^0_{i}$ at the beginning of the recombination region; these are evaluated as the maxima of \radicals, or the values of $X^0_{i}$ at the location of the maximum of $X_{\ce{H}}$ in the case of \ce{H2} and \ce{O2}. $X_{\ce{O2}}$ is plotted separately for clarity.
	}
	\label{H2RadicalsSpecies}
\end{figure}

This investigation is relevant to the structure of the OH* -- excited OH -- chemiluminescence radiation, studied by Graña and Mahmoudi in \cite{GRANAOTERO2019115750} where it is shown that a significant fraction of the total radiated intensity can be originated and supported by the underlying \ce{H2} flame recombination region. Similarly, the recombination region plays an important role in the \nox distribution structure in \ce{H2} flames. More specifically, as will be shown in coming work, the persistent concentrations of NO and \ce{N2O} left behind the fuel consumption layer are actually supported by the underlying recombination region studied in this paper.

The rest of the paper starts with Section \ref{RecombinationRegion}. {\it The recombination region}, giving an overview of the main properties of the recombination region. Section {\it{\ref{SimplifiedMechanism}. Kinetics in the recombination region}}, analyzes in detail its kinetic structure using a reduced version of the \ce{H2}-air mechanism which is simple but still contains the essential elements. Section \ref{Formulation}. {\it{Formulation}}, studies in detail the problem of the structure of the recombination region, showing how it can be simplified and solved. 
Finally, Section \ref{Conclusions}. {\it{Discussion and conclusions}}, discusses the limitations and applicability of the findings. In particular, the effect of heat losses on the structure of the recombination region is addressed in some detail to determine when, and to some extent how, the analysis should be modified in their presence. 

\section{The recombination region.} \label{RecombinationRegion}

Chemical equilibrium, or equivalently zero gradients of the temperature and species distributions, are the appropriate boundary conditions downstream of a planar, steady premixed flame. These are to be applied past the flame, at a distance large enough to ensure that thermodynamic equilibrium has been attained. However, in analytical studies, they are often applied immediately downstream of the fuel consumption layer, as if equilibrium were reached right after it. The predictions so obtained are in good agreement with the numerical or experimental results. Nevertheless, when solving numerically (exactly) this problem, it is found that these conditions need to be enforced at a much larger distance downstream of the fuel consumption layer in order to avoid unphysical discontinuities in the gradients.

Figure \ref{H2RadicalsSpecies} shows an example of such numerical solutions in the case of a \ce{H2}-air flame with an equivalence ratio $\phi = 0.6$. These results and those of the rest of the paper have been obtained with a customized version of the open source code Cantera \cite{cantera}, together with the Glarborg et al. \cite{GLARBORG201831} kinetic mechanism with inert \ce{N2}. The results obtained with the full (including the \ce{N2} oxidation) mechanism are indistinguishable from the results presented below. Similar results, with inessential differences in the numerical results, are obtained as well with the San Diego \cite{SanDiego} or the more recent Konnov \cite{KONNOV201914} mechanisms. 

As can be seen, the temperature, \ce{O2} and \ce{H2O} rapidly reach their final equilibrium concentrations, so the zero-gradient boundary conditions seem accurate for them right after the fuel consumption layer. However, the concentrations of \ce{H2} and radicals slowly decay over a relatively large length until they finally reach the zero-gradient condition at the final burnt equilibrium state. 

Understanding this behavior and in particular the length scale associated with the radicals decay requires the analysis of the interplay between transport and kinetics, which are briefly overviewed in the rest of this section, also anticipating the main results analyzed in detail in the rest of the paper.

\begin{figure}[h!]
	\begin{center}
	\includegraphics[width=0.44\textwidth]{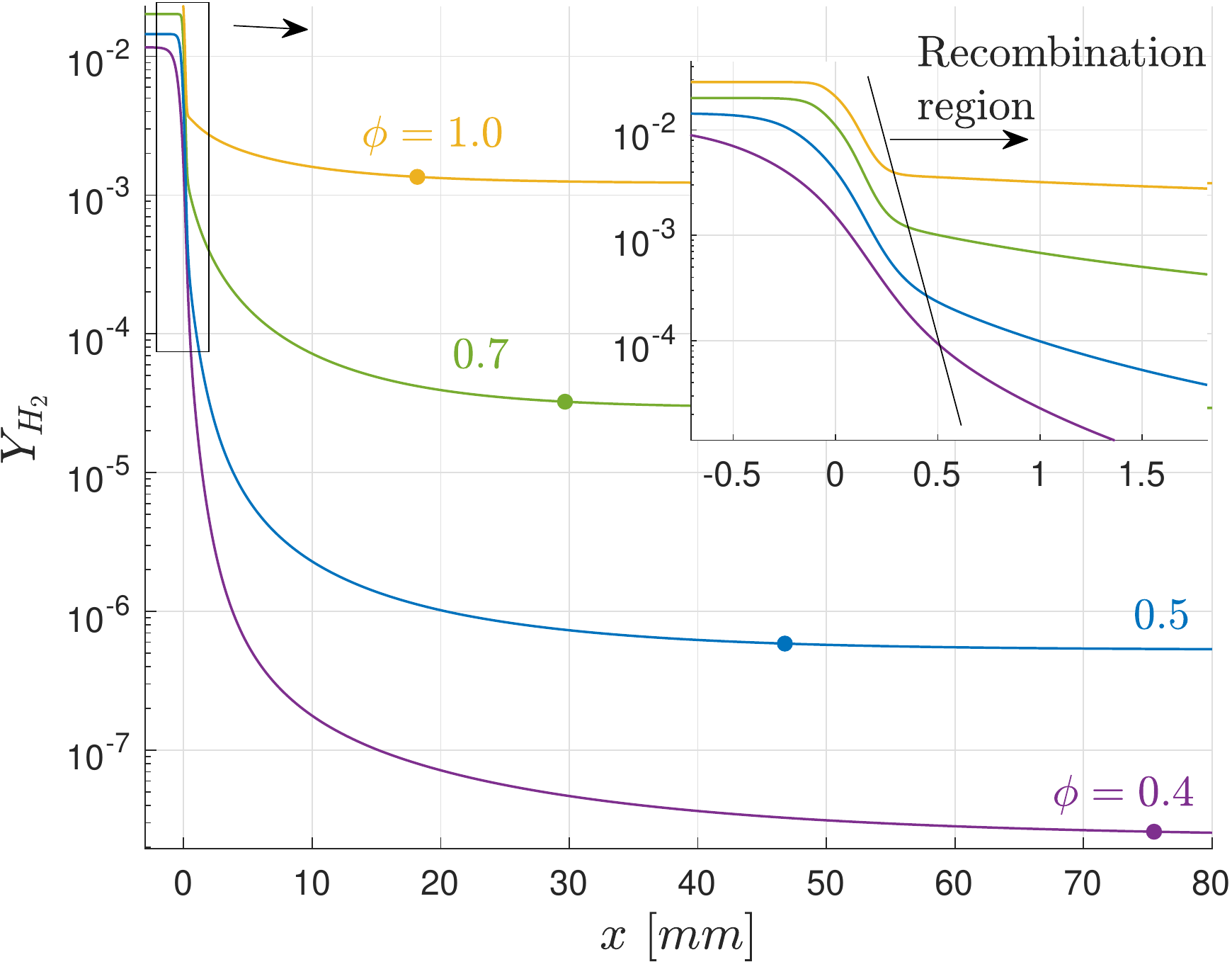} \hspace{1cm}
	\includegraphics[width=0.44\textwidth]{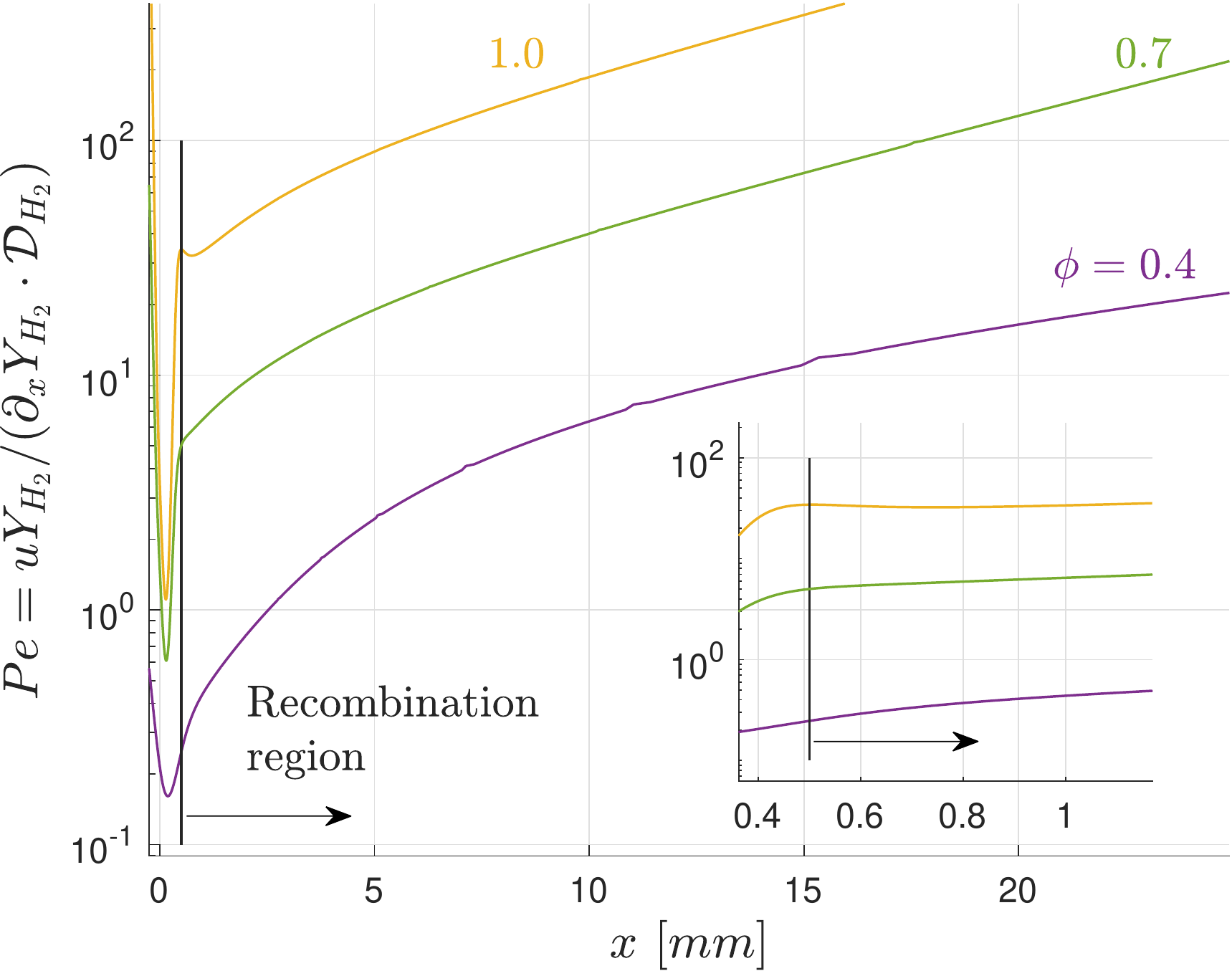}
	\caption{The left plot represents the \ce{H2} mass fraction distributions for several representative values of the equivalence ratio. The dots mark the location where $Y_{\ce{H2}}$ has decayed to $Y^b_{\ce{H2}}/Y_{\ce{H2}} = 0.9$ with $Y^b_{\ce{H2}}$ the final burnt equilibrium mass fraction. 
	The right plot represents the corresponding distributions of the P\'eclet number $Pe_r = u\mathcal{L}/\mathcal{D}_{\ce{H2}}$, based on $\mathcal{L}$ and on the \ce{H2} mass diffusivity. This plot shows that the recombination region is essentially a convective-reactive region, with $Pe_r \gg 1$, for lean flames with not to small values of $\phi$. }
	\label{LengthsAndPeclets}
	\end{center}
\end{figure}

\subsection{Transport.}  \label{Transport}

The reduced but finite reaction rates prevalent in the recombination region make its length scale larger than, but comparable to, that of the preheat region, as suggested by Figure \ref{H2RadicalsSpecies}. 
As a result of this length scale disparity, the gradients in the recombination region are much smaller than those in the fuel consumption layer; they can therefore be set to zero -- chemical equilibrium in a first approximation -- when the solution is seen with the scale of the fuel consumption layer. Accordingly, radicals concentrations remain constant downstream of the fuel consumption layer at distances comparable to its length scale, as in the top plot of Figure \ref{H2RadicalsSpecies}. 
However, when the proper scale for the much longer recombination region is used, as in the middle plot of this Figure, concentrations are seen to slowly decay and finally level off to zero slopes -- chemical equilibrium to all orders.

\begin{figure}[h!]
	\begin{center}
		\includegraphics[width=0.44\textwidth]{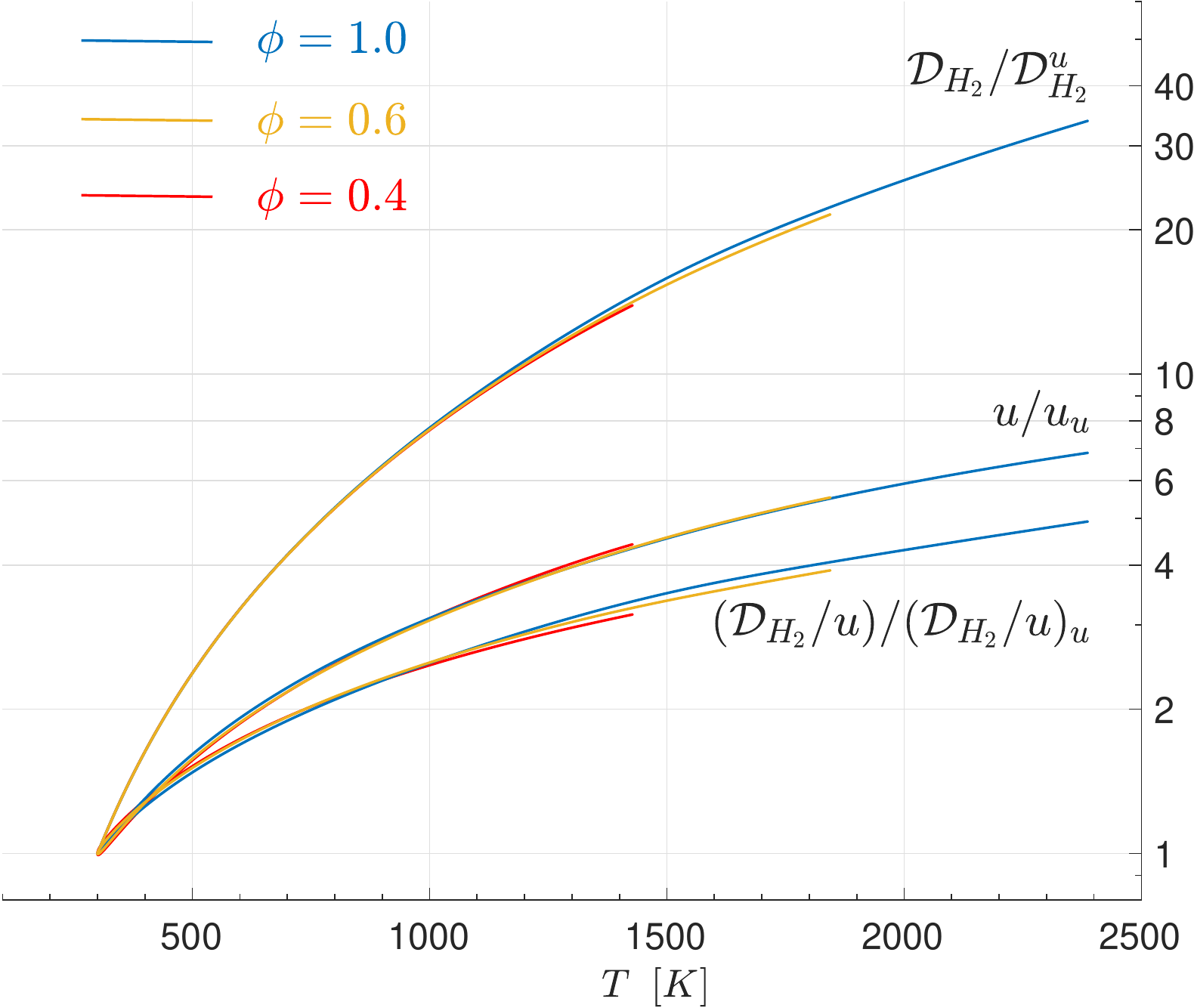} 
	\caption{Distributions in terms of the temperature of relevant physico-chemical magnitudes, showing in particular that despite the moderately large variations of the \ce{H2} diffusion coefficient $\mathcal{D}_{\ce{H2}}$, the length scale $\mathcal{D}_{\ce{H2}}/u$ changes at most by a factor of 5 between the cold and hot sections of the flame.}
	\label{LengthPreheatAcrossFig}
	\end{center}
\end{figure}

This characteristic length -- larger even than that of the preheat region where convective and diffusive transport are of the same order of magnitude -- render the transport of species and enthalpy dominated by convection in the first approximation. A quantitative measure of the relative magnitude of each transport mode, shown in Figure \ref{LengthsAndPeclets}, is the P\'eclet number $Pe_r = u\mathcal{L}/\mathcal{D}_{\ce{H2}}$, based on the local characteristic length $\mathcal{L} = Y_{\ce{H2}}/(\partial Y_{\ce{H2}}/\partial x)$. $Pe_r$ is also of the order of the ratio $\mathcal{L}/\delta_L$ of $\mathcal{L}$ to the flame thickness (without the recombination region) $\delta_L \sim \mathcal{D}_{\ce{H2}}/u$, based on the \ce{H2} mass diffusivity and on the local flow velocity $u$ (in a frame of reference attached to the flame), of the order of the burnt products flame speed\footnote{The preheat region length $\delta_L$ should be estimated using the values of $\mathcal{D}_{\ce{H2}}$ and $u$ upstream the fuel consumption layer (at the fresh mixture temperature for instance), whereas those entering in the definition of $Pe_r$ are evaluated in the recombination region, around the burned equilibrium temperature. However, even in the hottest flames near stoichiometry, the length $\delta_L$ remains essentially of the same order of magnitude on both sides of the flame (see Figure \ref{LengthPreheatAcrossFig}).}.

The recombination region is thus characterized by moderately larger than unity Péclet (or Reynolds) numbers. Therefore, the transport problem for the distribution of species and temperature transitions from the diffusion-controlled regime of the fuel consumption layer to the convection-controlled regime of the recombination region. Or in mathematical terms, from the elliptic character of the transport problem in the fuel consumption layer to a parabolic character in the recombination layer. Once in the recombination region, information is only transported downstream, away from the fuel consumption layer. 

The recombination region structure is thus slave of the conditions left past the fuel consumption layer, which are in turn independent of the evolution in the recombination region. This explains why the condition of chemical equilibrium applied on the hot edge of the fuel consumption layer works well. The recombination region can thus be ignored, in the first approximation, when the interest is in calculating the flame velocity.

\begin{figure}[h!]
	\begin{center}
		\includegraphics[width=0.44\textwidth]{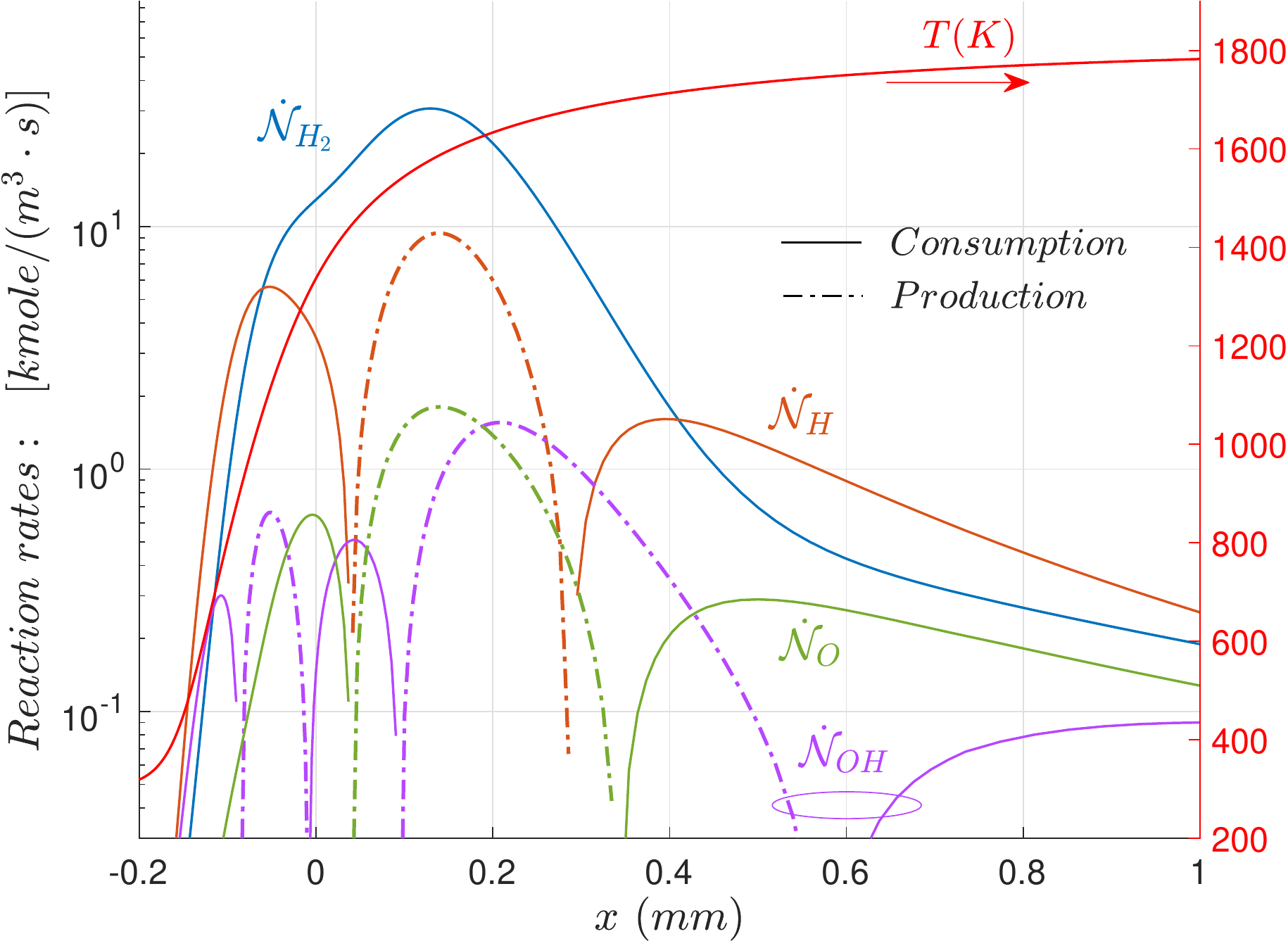} \hspace{1cm}
		\includegraphics[width=0.44\textwidth]{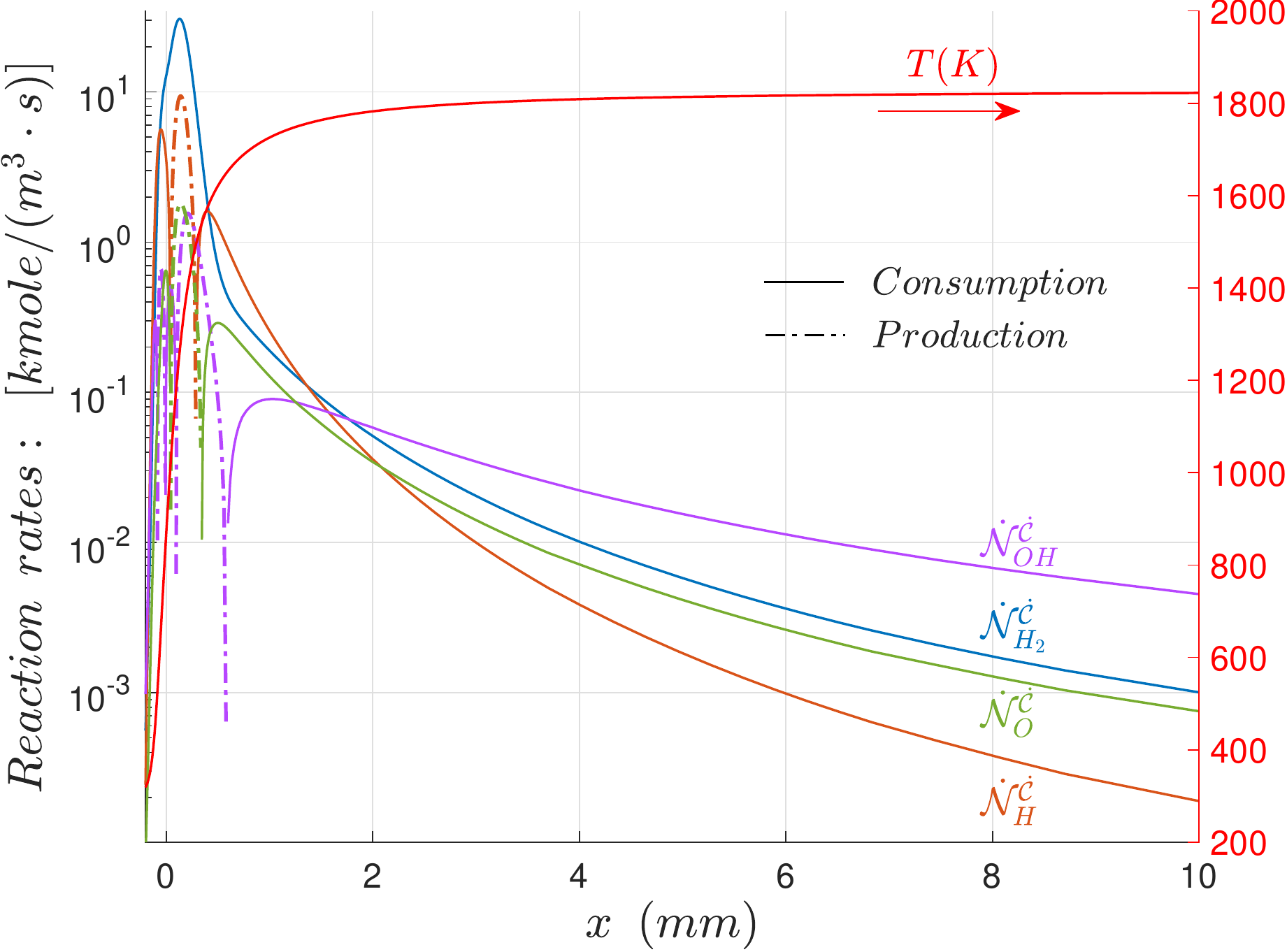}
	\end{center}	
	\caption{Distributions of the net $\dot{\mathcal{N}}$ consumption/production (solid/dash-dot lines) rates of \ce{H2}, O, H and OH in the fuel consumption layer (top) and downstream of it in the recombination region (bottom). The transition between these two regions can be identified by the abrupt drop of two orders of magnitude of the \ce{H2} consumption rate and also by the transition from net production to net consumption of the three radicals O, H and OH.}
	\label{NetRates}
\end{figure}
\subsection{Kinetics.} The distributions of the net consumption rates of radicals and \ce{H2} of Figure \ref{NetRates} show that, although at reduced rates compared with those in the fuel consumption layer, the flame is still reacting downstream of the fuel consumption layer. These plots also illustrate the change of sign of the net production rates of radicals which transition from net production in the fuel consumption layer to net consumption in the recombination region. They clearly show thus, as anticipated, the role of this region depleting, down to the final thermodynamic equilibrium state, the unburned \ce{H2} and the excess of radicals left past the fuel consumption layer. 

\begin{table}[h!]
	\begin{center}
	\begin{tabular}{l r @{} c @{} l }
			Label & & Reaction &  \\
			\midrule 
			 \multicolumn{2}{l}{\small \hspace{.1cm} shuffle reactions}  &   & \\
			\cmidrule(l){1-2}
			1 &  \ce{O2} + \ce{H} &$\lra$ & \ce{O} + \ce{OH}  \\
			4 &  \ce{H2} +  \ce{OH} &$\lra$ & \ce{H2O} + \ce{H}  \\
			5 &  \ce{2 OH}  &$\lra $ & \ce{H2O} + \ce{O}  \\[.05cm]
			 \multicolumn{2}{l}{\small \hspace{.1cm} \ce{HO2} reactions}  &   & \\
			\cmidrule(l){1-2}
			13 &  \ce{H} +  \ce{O2} +  \ce{M} &$\lra$ &  \ce{HO2} + \ce{M}  \\
			17 &  \ce{HO2} +  \ce{O} &$\lra$ &  \ce{O2} + \ce{OH}  \\
			19 &  \ce{HO2} +  \ce{OH} &$\lra$ & \ce{H2O} + \ce{O2}  \\
			\midrule \\
	 \end{tabular}
	\caption{Reduced \ce{H2}-\ce{O2} kinetic mechanism in the recombination region. This mechanism reproduce the exact (that is, the numerical results with the full mechanism) net production rates of all the species with an accuracy of at least 10\% in the recombination region of lean flames. The first three, the shuffle reactions, are out of equilibrium in the fuel consumption layer but remain in nearly exact partial equilibrium throughout the recombination region. The last three reactions represent the basic recombination engine recycling radicals and \ce{H2} to products. Although these three reactions are early in the recombination region out of equilibrium (see Figure \ref{NetRates}), this subset is considered as reversible in order to capture the final thermodynamic equilibrium, where all reactions eventually reach partial equilibrium. The reaction numbers refer to Glarborg et al.'s mechanism \cite{GLARBORG201831}, excerpted in \ref{FulH2O2Mechanism}
	.} 
	\label{KineticMechanism}
	\end{center}
\end{table}

\begin{figure}[h!]
	\begin{center}
		\includegraphics[width=0.44\textwidth]{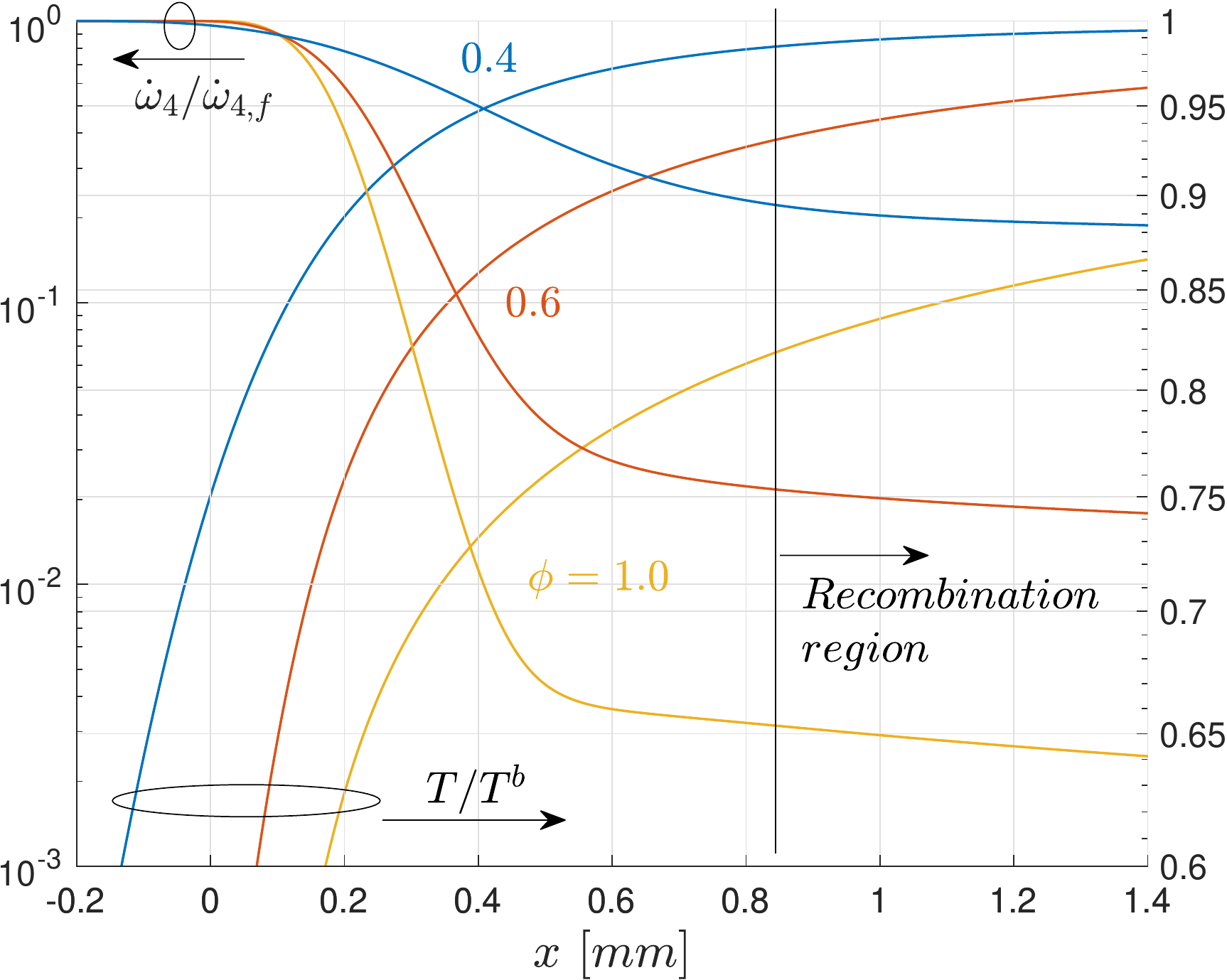} \hspace{1cm}
		\includegraphics[width=0.44\textwidth]{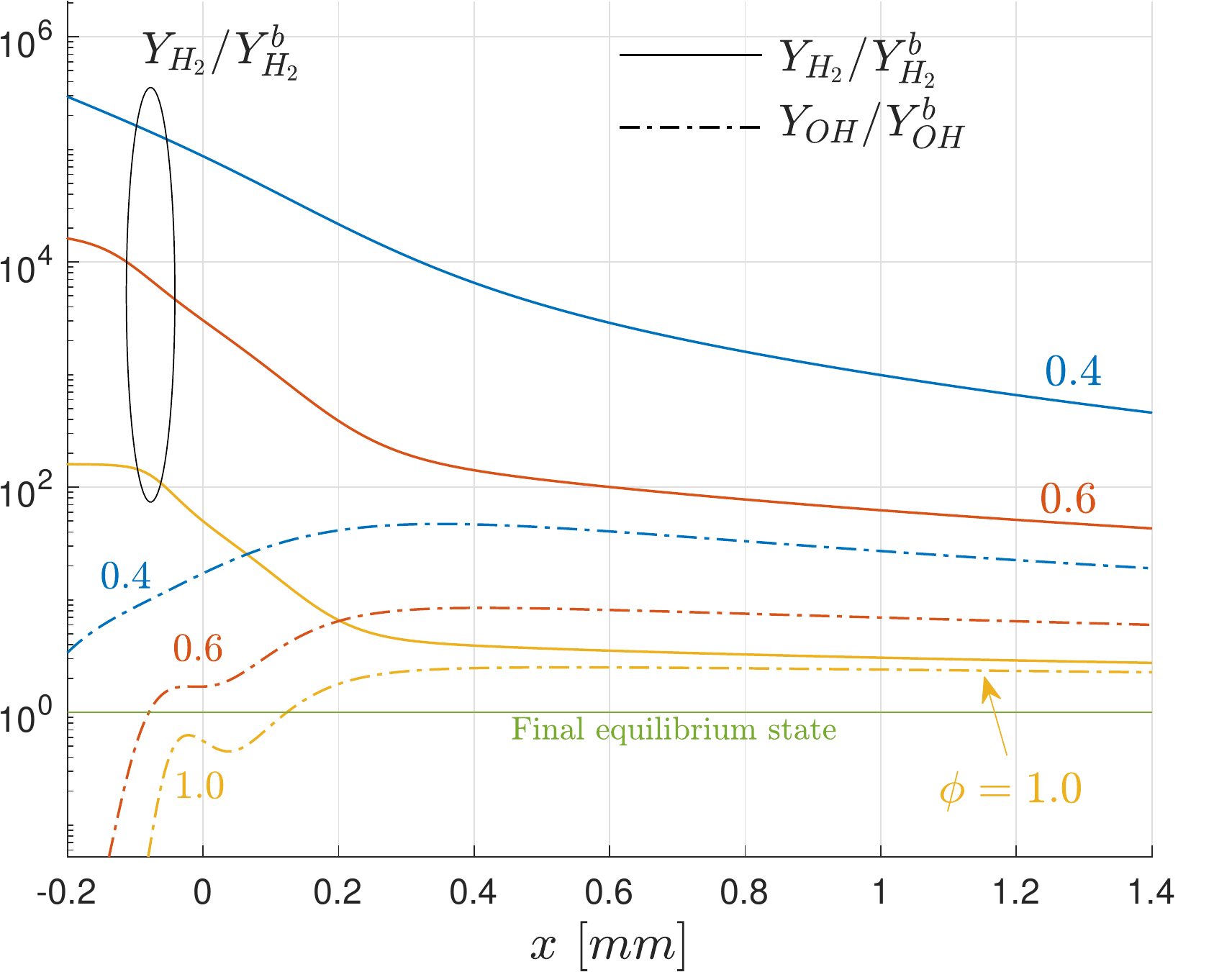}
	\end{center}	
	\caption{The left plot illustrates, for several representative lean flames, the transition between the fuel consumption layer (upstream) and the recombination region (downstream) using as indicator the net reaction rate of the shuffle reaction 4, \ce{H2} + \ce{OH} $\lra$ \ce{H2O} + \ce{H}. The transition can be identified by the drop, sharper and deeper as stoichiometry is approached, to values much smaller than unity of $\dot{\omega}_4/\dot{\omega}_{4, f}$. As shown in the right plot, this drop -- quench of reaction 4 -- leaves past the fuel consumption layer values of the mass fraction $Y_{H_2}$ that are still orders of magnitude larger than the final equilibrium values $Y^b_{H_2}$. The temperature $T/T^b$, scaled with its final value $T^b$, is also shown for reference. Notice that all the magnitudes displayed in the right plot have unity final equilibrium values, showing that, especially in low-$\phi$ flames, the recombination region is characterized by a many orders of magnitude decay of the scaled mass fraction $Y_{H_2}/Y^b_{H_2}$. Note also that close to stoichiometry, a significant amount of the heat release occur downstream of the fuel consumption layer, making inaccurate the assumption of an isothermal recombination region. The vertical line marking the beginning of the recombination region is suggestive, not a sharp transition.}
	\label{Transition}
\end{figure}

This reactivity is ultimately associated with the irreversibility of the fuel consumption layer, where the fast, out-of-equilibrium conversion of \ce{H2} to products involves radicals in amounts well in excess over the values corresponding to chemical equilibrium. However, the reactions responsible for the \ce{H2} fast consumption -- the shuffle reactions in Table \ref{KineticMechanism} -- abruptly quench at the hot edge of the fuel consumption layer, leaving \ce{H2} and the radicals \radicals frozen in the first approximation, unable to decay to equilibrium. This transition to partial equilibrium (quench) of the shuffle reactions, illustrated in Figure \ref{Transition} in the specific case of the reaction 4, marks the end of the fuel consumption layer and the beginning of the recombination region.

The almost exact partial equilibrium of the three shuffle reactions, maintained throughout the entire recombination region, gives three expressions (see \eqref{PartialEquilibriumSolvedOnlyH2} below) yielding the radicals \radicals concentrations in terms of that of \ce{H2}. These expressions replace the corresponding transport equations, hence reducing the number of degrees of freedom by three. If, in addition, the temperature and \ce{O2} and \ce{H2O} concentrations are assumed to be constants, as suggested by Figure \ref{H2RadicalsSpecies}, the full flame problem can be reduced to just solving the \ce{H2} distribution, given by the ordinary differential equation derived below in \eqref{H2FinalEqDimensionless}.

The source term in this equation is due to the \ce{HO2} kinetics, which drives the chemical activity in the recombination region. These reactions had in the fuel consumption layer much smaller rates than those of the shuffle reactions. In the recombination region however, they remain out of equilibrium, i.e. with widely different forward and reverse rates, and become dominant, setting in particular the scale for the deviations from partial equilibrium of the shuffle reactions.

In summary, the realistic description of \ce{H2} flames with the detailed kinetic mechanism can be reduced, in the recombination region, to a one-degree-of-freedom, thick, isothermal, diffusion-free structure trailing the flame, similar to that found by Li\~n\'an \cite{LinanRecombination} as the slow recombination regime in his analysis of \ce{H2}-air flames, based on a simplified, two-step model due to Zel'dovich \cite{Zeldovich1961TwoSteps}. 



\section{Kinetics in the recombination region.} \label{SimplifiedMechanism}


The basic recombination working principle can be exemplified with the reduced mechanism of Table \ref{KineticMechanism}, which reproduces the net production and consumption rates with errors of the order of 10\%. The characteristic scales and the fundamental mathematical structure found with it remain the same when the full mechanism is used. 

\begin{figure}[h!]
	\begin{center}
		\includegraphics[width=0.44\textwidth]{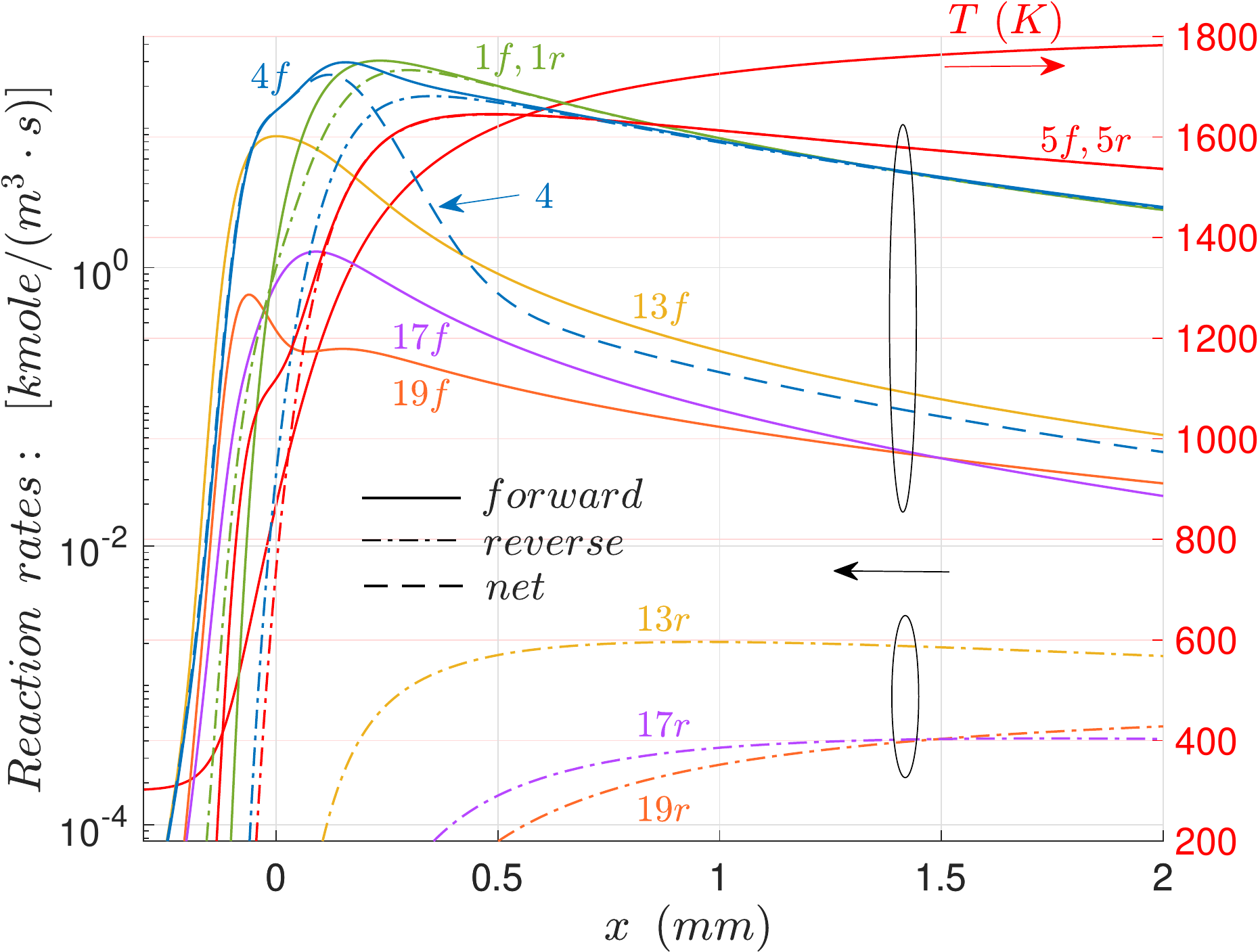} \hspace{1cm}
		\includegraphics[width=0.44\textwidth]{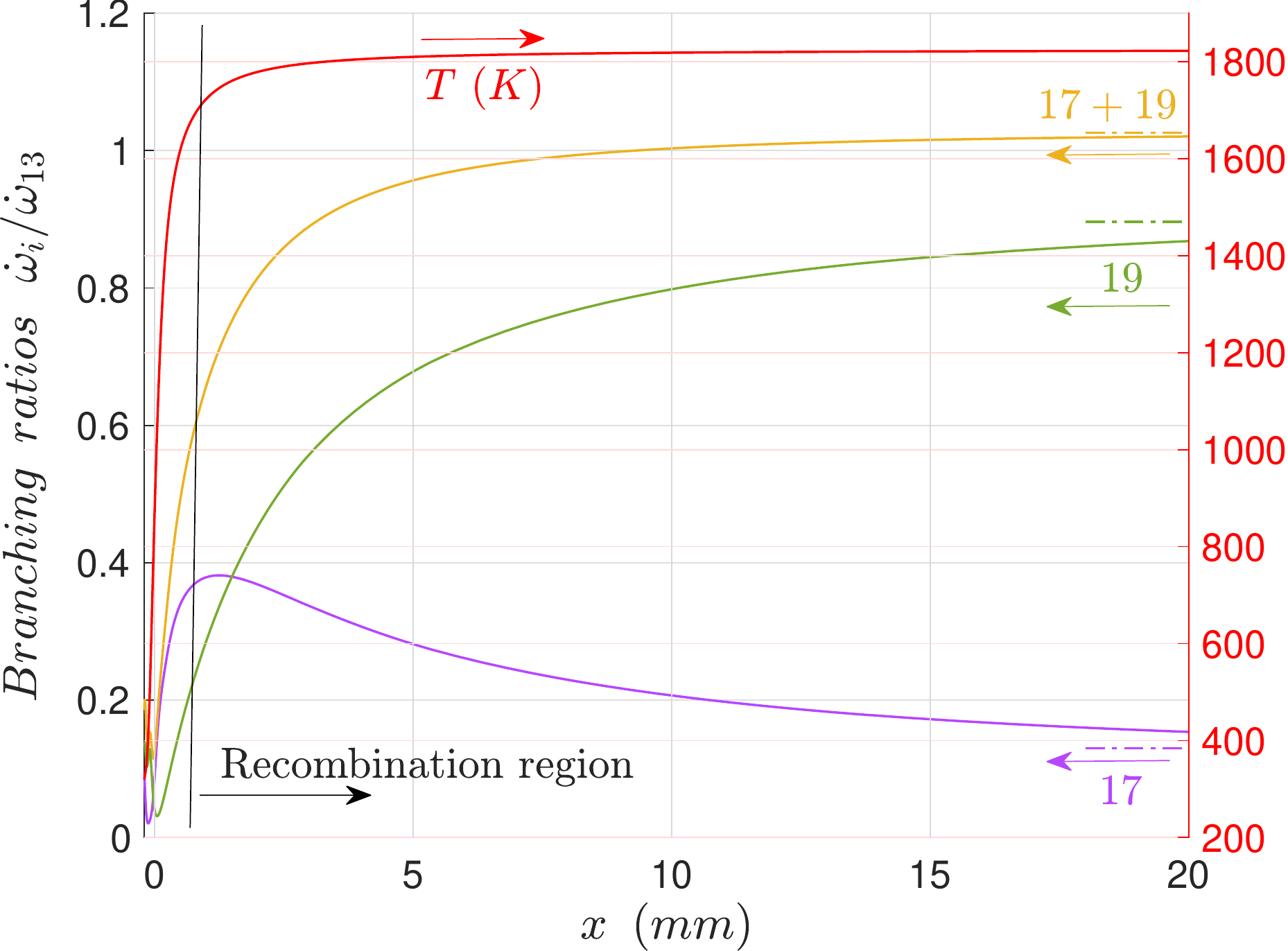}
	\end{center}
	\caption{ Kinetics in the fuel consumption layer, showing that the shuffle reactions (1, 4 and 5) are out of equilibrium throughout it but converge to partial equilibrium on its hot side. The kinetics in the recombination region is instead characterized by the almost exact partial equilibrium (balanced forward (solid) and reverse (dash-dot) rates) of the shuffle reactions; in contrast, the reactions of the \ce{HO2} kinetics are clearly out of equilibrium, with much faster forward than reverse rates. Note as well that the deviations from partial equilibrium of the shuffle reactions (for clarity only the net rate (dash line) of reaction 4 is plotted) are of the order of the forward rates of the \ce{HO2} reactions. The bottom plot shows the distribution of the branching ratio $\alpha = \dot{\omega}_{17} /\dot{\omega}_{13}$ and $1 - \alpha = \dot{\omega}_{19} /\dot{\omega}_{13}$ together with their combined net rate (yellow line). Deviations of this line from unity, notably early in the recombination region, indicate inaccuracies of the reduced mechanism. The vertical line suggesting the origin $x \approx 0.5$ mm of the recombination is guided by the plot on the left.}
	\label{DifferentKinetics}
\end{figure}

This mechanism comprises the three shuffle reactions and the three most relevant reactions from the full \ce{HO2} kinetics, excerpted in \ref{FulH2O2Mechanism}
. This last subset represents the engine driving the conversion of the radicals and \ce{H2} excess into products. As shown in Figure \ref{DifferentKinetics}, these reactions are left past the fuel consumption layer out of equilibrium, with much larger forward than reverse rates, only reaching partial equilibrium asymptotically, as the final burnt equilibrium state is approached. 

In contrast, the shuffle reactions are, in the first approximation, in partial equilibrium, with balanced forward and reverse rates throughout the entire recombination layer. However, the deviations of the shuffle reactions from the exact partial equilibrium still need to be considered. This can be easily seen from the kinetics of \ce{H2}, whose consumption rate $\dot{\mathcal{N}}_{\ce{H2}}$ is simply given by the net rate of shuffle reaction 4 (see also Figure \ref{H2Kinetics} below), which is non-zero only in the approximation beyond partial equilibrium. These deviations from exact partial equilibrium are in turn, as will be shown, of the order of the rates of the \ce{HO2} kinetics $\dot{\mathcal{N}}_{\ce{H2}} \sim \dot{\omega}_{4} \sim \dot{\omega}_{13}$.



\subsection{The recombination mechanism.} The distributions of the rates of the \ce{HO2} reactions are represented in Figure \ref{DifferentKinetics}, which shows that, early in the recombination region, the forward rates of reactions 13, 17 and 19 are several orders of magnitude faster than the reverse ones. The kinetics of \ce{HO2} can thus be seen as a two-step mechanism. First, \ce{HO2} is produced through reaction 13f: \ce{H} + \ce{O2} + \ce{M} $\rightarrow$ \ce{HO2} + \ce{M}. This is the rate limiting step in the \ce{HO2} kinetics and therefore sets the characteristic chemical time in this region. Balancing its production rate, the chain-breaking reactions 17 and 19 consume \ce{HO2} and radicals converting them into \ce{O2} and \ce{H2O}. This basic scheme, namely reaction 13 producing \ce{HO2} and the rest of the reactions consuming and maintaining it in steady state, remains unmodified when the full \ce{HO2} mechanism is considered.

The steady state of \ce{HO2} can be written as $\dot{\omega}_{13} \approx \dot{\omega}_{17} + \dot{\omega}_{19}$, which shows that the mass of \ce{HO2} is conserved. \ce{HO2} serves thus just as the intermediate in the engine pumping radicals to products. 

The overall stoichiometry of the \ce{HO2} subset can be made more explicit using the branching ratio $\alpha = \dot{\omega}_{17} /\dot{\omega}_{13}$, so $1 - \alpha \approx \dot{\omega}_{19}/\dot{\omega}_{13}$, both plotted in Figure \ref{DifferentKinetics}. Using it to eliminate \ce{HO2} among the three reactions gives
\begin{gather}
	\label{OverallReaction}
	\ce{H} + \alpha \ce{O} + (1 - 2\alpha) \ce{OH} \rightarrow (1 - \alpha) \ce{H2O},
\end{gather}
which clearly shows that the overall effect is the conversion of radicals to \ce{H2O}. 

Note also the absence of \ce{O2} from this equation, showing that its mass is conserved, just as that of \ce{HO2}. In short, \ce{HO2} is produced from \ce{O2} through reaction 13 and recycled back to \ce{O2} through the recombination reactions 17 and 19.

The overall reaction \eqref{OverallReaction} shows that the \ce{HO2} submechanism represents net consumption rates -- sinks -- of radicals and the production rates -- sources -- of \ce{H2O}, which in turn induce corresponding net rates on the shuffle reactions. These are thus removed from partial equilibrium and forced to sustain net rates of the order of those of the \ce{HO2} reactions, i.e. $\dot{\omega}_{1,4,5} \sim \dot{\omega}_{13}$. Intuitively, -- or using Le Chatelier's principle -- it can be anticipated that the deviations are such as to promote the relaxation back to partial equilibrium; that is, they are faster in the chain-breaking direction, which converts the excess of radicals and \ce{H2} into \ce{O2} and \ce{H2O}. For instance, reaction 4 is slightly faster in the forward than in the reverse direction, so it produces one molecule of \ce{H2O} per each consumed molecule of \ce{H2}. 

\begin{figure}[h!]
	\begin{center}
	\includegraphics[width=0.45\textwidth]{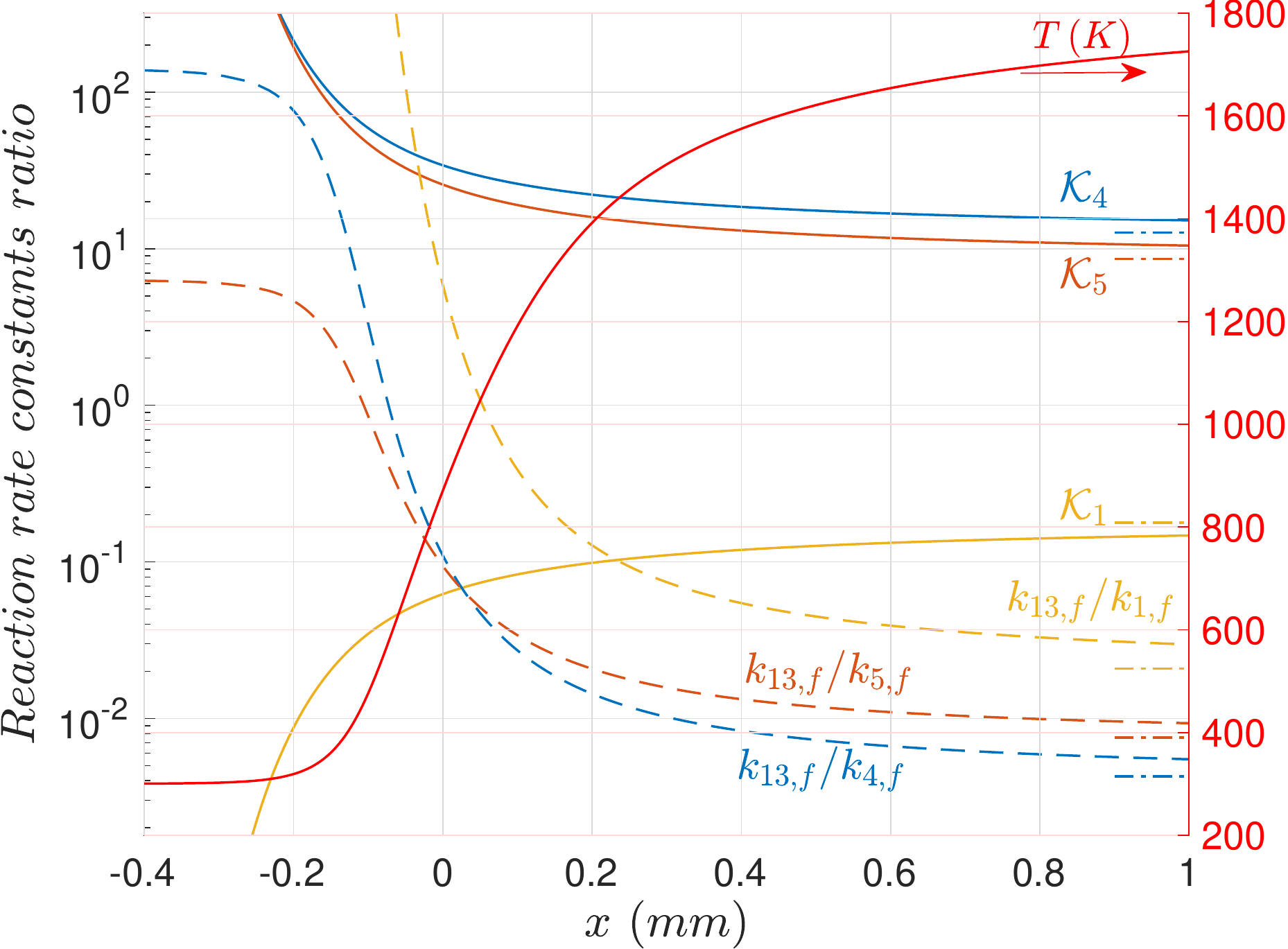} \\[0.5cm]
	\caption{Distributions, in a $\phi = 0.6$ \ce{H2}-air flame, of the shuffle reactions equilibrium constants $\mathcal{K}_{1,4,5}$, which reach their final high-temperature asymptotic values quickly downstream of the hot edge of the fuel consumption layer, so they are nearly constants throughout the recombination region. Also shown in dash lines the ratios $k_{13,f}/k_{1, 4, 5;f} \sim 10^{-2}$. The short dash-dot segments represent the burnt equilibrium values.}
	\label{EquilibriumConstants}
	\end{center}
\end{figure}

More quantitatively, Figure \ref{DifferentKinetics} shows that the net rates of the shuffle reactions are two orders of magnitude smaller that their corresponding forward or reverse rates; that is, the deviations (the net rates) from exact partial equilibrium are small: $\dot{\omega}_{i, f} - \dot{\omega}_{i, r} \sim \dot{\omega}_{13} \sim 10^{-2}\dot{\omega}_{i, f} \sim 10^{-2}\dot{\omega}_{i, r}$ for $i = 1, 4, 5$. Ultimately, the factor of two orders of magnitude separating these rates is determined by the ratios $k_{13,f}/k_{1, 4, 5;f} \sim 10^{-2} $ of the reaction constants of reaction 13 to those of the shuffle reactions, as shown in Figure \ref{EquilibriumConstants}. In the multiple-scales and perturbation methods terminology, $\epsilon_{1,4,5} = k_{13,f}/k_{1, 4, 5;f}$ are the small parameters causing the phenomena described in this paper.


%

\subsection{Partial equilibrium of the shuffle reactions.}

Although well-known, the partial equilibrium of the shuffle reactions is briefly discussed here to introduce some nomenclature that will be used hereafter. Balancing their forward and reverse rates yields
\begin{subequations}
	\begin{gather}
		\C{O} \approx \mathcal{K}_{1}\mathcal{K}_{4}  \frac{\C{O2}}{\C{H2O}} \, \C{H2} , \\
		\C{OH} \approx \sqrt{ \frac{\mathcal{K}_{1}\mathcal{K}_{4}}{\mathcal{K}_{5} } \, \C{O2} \C{H2} }  , \\
		\C{H} \approx \sqrt{ \frac{\mathcal{K}_{1}\mathcal{K}_{4}^3} { \mathcal{K}_{5} } \frac{\C{O2}} {{\C{H2O}}^2}  \, \C{H2}^3},
	\end{gather}%
	\label{PartialEquilibriumSolved}%
\end{subequations}
with the equilibrium constants $\mathcal{K}_i = k_{i, f}/k_{i, r}$, functions of only the temperature. The relative errors $|\C{S}_{exact}-\C{S}_{approx}|/\C{S}_{exact}$ (S = \radicals) of these approximations are of the order of the relative deviations $\mathcal{E}_i = (\dot{\omega}_{i, f} - \dot{\omega}_{i, r})/\dot{\omega}_{i, f}$ from partial equilibrium of the shuffle reactions (see Graña and Mahmoudi \cite{GRANAOTERO2019115750}), which, for not too lean flames, are typically of order $\mathcal{E}_i \sim 10^{-2}$ (see Section \ref{Conclusions}. {\it{Discussion and conclusions.}} and Figure \ref{PartialEquilibrium} for more details).

Guided by Figures \ref{H2RadicalsSpecies} and \ref{ErrorsOnlyH2}, which show that the temperature soon reaches its final equilibrium value, the equilibrium constants $\mathcal{K}_{i}$ can be assumed, throughout the recombination region, to be constants equal to their final burnt equilibrium values at $T^b$. Similarly, Figure \ref{H2RadicalsSpecies} 
suggests that the concentrations of \ce{O2} and \ce{H2O} can as well be replaced in these expressions by the corresponding burnt equilibrium values. 

With these approximations, the equations \eqref{PartialEquilibriumSolved} can be more compactly written using the burnt equilibrium concentrations $\C{O}^b$, $\C{H}^b$ and $\C{OH}^b$ as
\begin{gather}
	\label{PartialEquilibriumSolvedOnlyH2}%
	\frac{\C{\ce{O}}^{e}}{\C{\ce{O}}^{b}} 
	= \left(\frac{\C{\ce{OH}}^{e}}{\C{\ce{OH}}^{b}}\right)^2 
	= \left(\frac{\C{\ce{H}}^{e}}{\C{\ce{H}}^{b}}\right)^{2/3}
	= \frac{\C{\ce{H2}}}{\C{\ce{H2}}^{b}}
	= \frac{1}{z},
\end{gather}
introducing the reduced \ce{H2} concentration $z = \C{\ce{H2}}^b/\C{\ce{H2}} \le1 $. The superscript $e$ remarks the use of the burnt equilibrium values of the equilibrium constants and \ce{O_2} and \ce{H2O} concentrations.

Despite being excellent approximations, these expressions are not yet enough because the concentration of \ce{H2} is still undetermined. They anticipate however that, in lean flames, the \ce{H2} concentration is the only degree of freedom in the recombination region. 
\begin{figure}[h!]
	\begin{center}
		\includegraphics[width=0.44\textwidth]{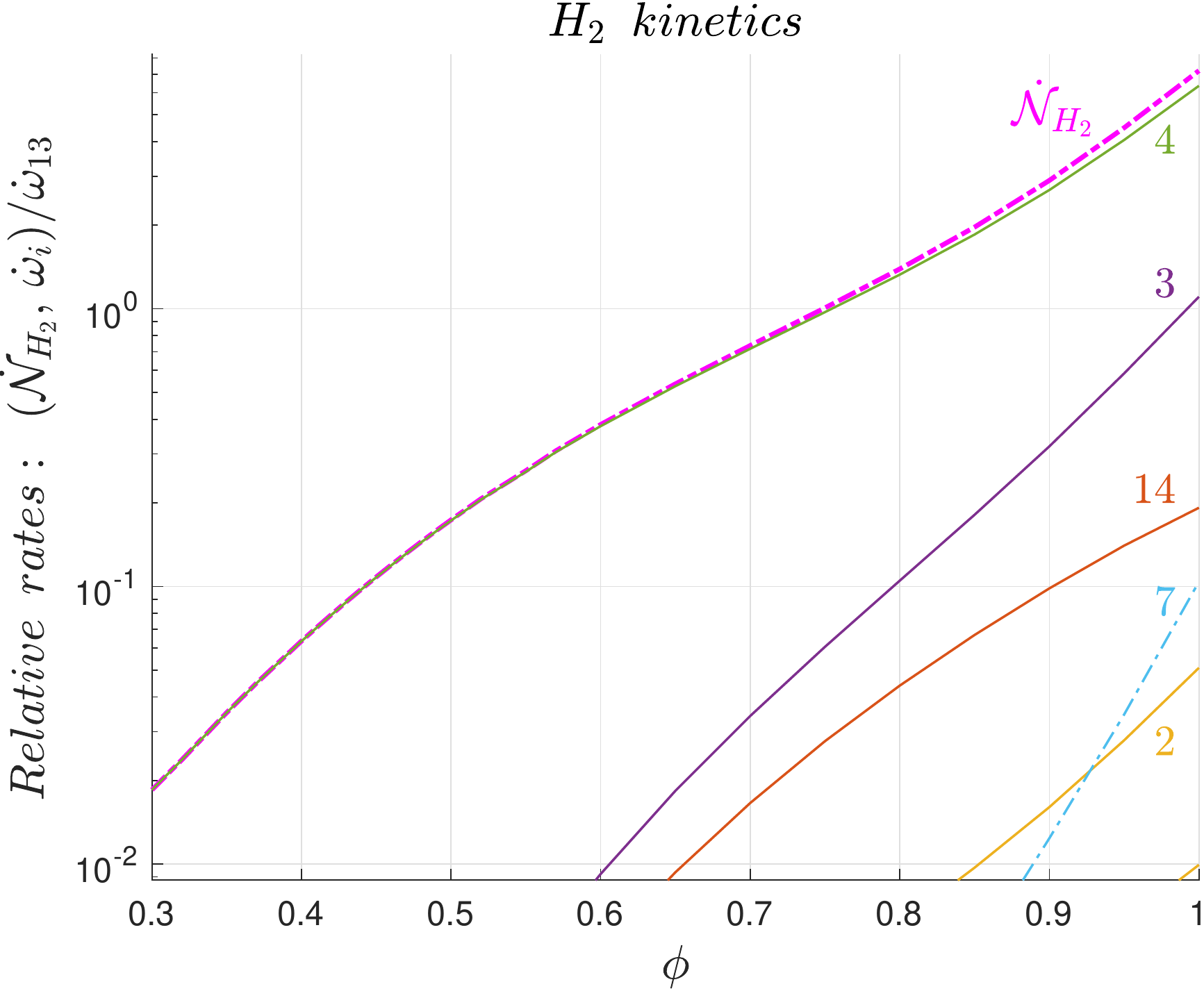} 
	\caption{Variation with the equivalence ratio $\phi$ of the net \ce{H2} consumption rate $\dot{\mathcal{N}}_{\ce{H2}}$ and of the most contributing reactions $\dot{\omega}_i$, showing that, almost exactly, $\dot{\mathcal{N}}_{\ce{H2}} = \dot{\omega}_{4}$. All the rates are scaled with the net rate $\dot{\omega}_{13}$ of reaction 13, showing that this is the proper scaling for the kinetics in this region. The rates are evaluated at $x = 20$ mm -- with $x = 0$ mm the location where \ce{HO2} attains its maximum rate; $x = 0$ is thus representative of the fuel consumption layer location, and also of the beginning of the recombination region given the length contrast between both regions. This plot thus shows the asymptotic ordering, which is also representative of the ordering closer to the fuel consumption layer.}
	\label{H2Kinetics}
	\end{center}
\end{figure}

\section{Formulation}\label{Formulation}

The structure of the recombination region is, in moderately lean flames, essentially determined by the dynamics of \ce{H2}. The characteristic length scale is thus given by the balance between the transport rate of \ce{H2} and its consumption rate through reaction 4. This balance is first examined below in orders of magnitude to estimate the characteristic scales which will then be used to formulate and solve the problem.

These order of magnitude estimations are carried out here because they are usually even more insightful than actually solving the resulting problem, whose results can be easily guessed once the physics of the problem is understood. They are also helpful cleaning the problem out from small unimportant effects, and help making sure that what remains is relevant. In addition, the estimations are often enough to solve engineering problems, when trends are more important than precise numerical results.

\subsection{Scales in the recombination region.} 

The \ce{H2} consumption rate $\dot{\mathcal{N}}_{\ce{H2}}$ is almost exactly given by $\dot{\mathcal{N}}_{\ce{H2}} = \dot{\omega}_4$ through the entire range of $\phi$ in lean flames as illustrated in Figure \ref{H2Kinetics}. This Figure shows also that $\dot{\omega}_{13}$ is the appropriate scaling. Thus, in the first approximation
\begin{gather}
	\label{ReactionRateScale}
	\dot{\mathcal{N}}_{\ce{H2}} \approx \dot{\omega}_4 \sim \dot{\omega}^{e}_{13,f} \approx \C{H}^{e}/\tau^b_{13,f}, 
\end{gather}
where \C{H} approximated with $\C{H}^{e} \propto \C{H2}^{3/2}$, given by \eqref{PartialEquilibriumSolvedOnlyH2}, which is based on the final equilibrium values of the temperature and \ce{O2} and \ce{H2O} concentrations, also used to evaluate the burnt equilibrium value of the reaction time $\tau^b_{13,f} = 1/(k_{13,f}\C{O_2})|^{b}$. 

\begin{figure}[h!]
	\begin{center}
	\includegraphics[width=0.44\textwidth]{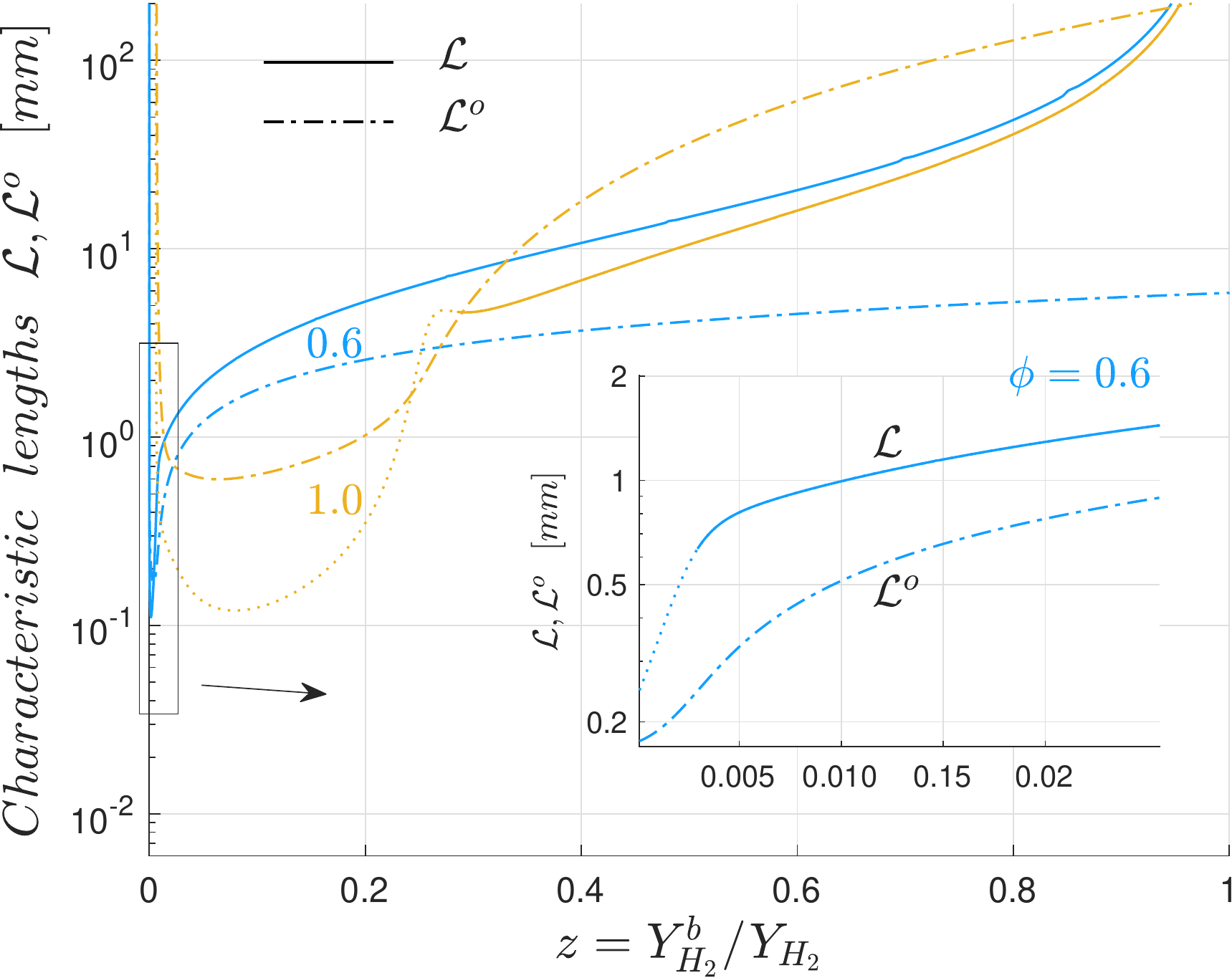} 
 	\caption{Comparison between $\mathcal{L} = Y_{\ce{H2}}/\partial_xY_{\ce{H2}}$ (solid lines) of Figure \ref{LengthsAndPeclets} and $\mathcal{L}^o$ (dash-dot lines) defined in \eqref{lengthScale}. $\mathcal{L}^o$ represents well the order of magnitude of $\mathcal{L}$ through the recombination region except, as expected, close to the final equilibrium, $z \rightarrow 1$, because only the forward rate of 13 is accounted for in $\mathcal{L}^o$. The dotted lines, included for reference, correspond to the fuel consumption layer, taken to roughly occur for $x < 0.6$ mm. }
	\label{Lengths}
	\end{center}
\end{figure}

On the other hand, the previous discussion suggests that convection can be used to estimate the species transport rate, so the \ce{H2} mass conservation equation reads, in orders of magnitude, as
\begin{gather}
	\label{ScalingH2}
	\dot{m}''_L \frac{\partial Y_{\ce{H2}}} {\partial x} \sim \mathcal{M}_{\ce{H2}} \dot{\omega}_4 \sim \mathcal{M}_{\ce{H2}} \dot{\omega}^{e}_{13,f},
\end{gather}
with $\dot{m}''_L = \rho u$ the mass flow rate of mixture consumed by the flame per unit surface and time. 

The characteristic length $\partial x \sim \mathcal{L}^o$ along which the \ce{H2} mass fraction undergoes variations of the order of itself, $\partial Y_{\ce{H2}} \sim Y_{\ce{H2}}$, is thus, in terms of the mole fractions $X_i$ to absorb for simplicity the molar masses,
\begin{gather}
	\label{lengthScale}
	\mathcal{L}^o = \frac{\dot{m}''_L Y_{\ce{H2}}}{ \mathcal{M}_{\ce{H2}} \dot{\omega}^{e}_{13,f} } = 
	\mathcal{L}^{o,b} \frac{X_{\ce{H2}}} {X^{e}_{\ce{H}}} = \mathcal{L}^{o,b} \sqrt{z}, 
\end{gather}
with $\mathcal{L}^{o,b} = u^b\tau^b_{13,f}$ the final equilibrium value of $\mathcal{L}^{o}$, written in terms of the flame's products side velocity $u^b = \dot{m}''_L/\rho^b $, quickly reached past the fuel consumption layer.

$\mathcal{L}^o$ is thus the length required to see variations of the \ce{H2} concentration of the order of $\C{H2}$ moles per unit volume with a consumption rate of order $\C{H}^{e}/\tau^b_{13,f} \propto \C{H2}^{3/2}$ moles per unit volume and time, so, going down the recombination region, it takes longer to reduce \C{H2} because its consumption rate decreases faster than linearly with \C{H2}. 


Although considered below, the reverse rate of reaction 13 is not taken into account in \eqref{ReactionRateScale} for simplicity. Figure \ref{Lengths} shows however that $\mathcal{L}^o$ compares well with $\mathcal{L} = Y_{\ce{H2}}/\partial_x Y_{\ce{H2}}$, except, naturally, close to the final equilibrium ($z \rightarrow 1$).

\subsection{Dimensionless formulation.} \label{DimensionlessFormulation} The transport equations for the steady flame propagation \cite{LinanWilliams, WilliamsBook, ZeldovichBook} can thus be written in dimensionless form using the scaled coordinate along the flame $\partial\xi = \partial x/\mathcal{L}^o \sim 1$ and the scaled rates $\tilde{\omega}_i = \dot{\omega}_i/\dot{\omega}_{13,f} \sim 1$. Dropping in addition the diffusion terms leads to the following form of the equations for \ce{H2} and radicals
\begin{subequations}
	\label{MainEquationsDimensionless}
	\begin{gather}
		\label{H2ME} \frac{\partial \ln Y_{\ce{H2}}} {\partial \xi}  = - \tilde{\omega}_{4} , \\
		\label{OME} \mathcal{D} X_{\ce{O}} = \tilde{\omega}_{5} + \tilde{\omega}_{1} + \tilde{\mathscr{S}}^e_{\ce{O}} , \\
		\label{OHME}
		\mathcal{D} X_{\ce{OH}} = \tilde{\omega}_{1} - \tilde{\omega}_{4} - 2 \tilde{\omega}_{5} + \tilde{\mathscr{S}}^e_{\ce{OH}} , \\
		\label{HME} \mathcal{D} X_{\ce{H}} =  \tilde{\omega}_{4} - \tilde{\omega}_{1} + \tilde{\mathscr{S}}^e_{\ce{H}} , \\ 
		\label{HO2ME} \mathcal{D} X_{\ce{HO2}} = \tilde{\mathscr{S}}^e_{\ce{HO2}} ,%
	\end{gather}
\end{subequations}
with the operator $\mathcal{D} \varphi = X_{\ce{H2}}^{-1} \partial \varphi /\partial \xi$. This system must be complemented with expressions for the rates, discussed in \ref{HO2RatesApp} and \ref{ShuffleRatesApp}, and with appropriate initial conditions.

The equations for the temperature, \ce{O2} and \ce{H2O} are left out of \eqref{MainEquationsDimensionless} because, as discussed in \ref{TemperatureAndProducts}, they show that the corresponding variations, of order $\partial (T-T^b)/(T^b-T^u) \sim \partial X_{\ce{O2}} \sim \partial X_{\ce{H2O}} \sim X_{\ce{H2}} $, are (not too close to stoichiometry), as shown in Figure \ref{H2RadicalsSpecies} and in Figure \ref{ErrorsOnlyH2} below, much smaller than the respective values $(T-T^b)/(T^b-T^u)$, $X_{\ce{O2}}$ and $X_{\ce{H2O}}$, thus validating approximating them as constants, as done for instance deriving \eqref{PartialEquilibriumSolvedOnlyH2}.

The rates in \eqref{MainEquationsDimensionless} have been split into those collected in $\tilde{\mathscr{S}}^e_{i}$, which include sources of species $i$ due to reactions that are out of equilibrium, i.e. with widely different forward and reverse rates. These can therefore be evaluated in the first approximation using \eqref{PartialEquilibriumSolvedOnlyH2} and expressed as functions of only $z$, as done in \ref{HO2RatesApp} for the \ce{HO2} reactions.

\begin{figure}[h!]
	\begin{center}
		\includegraphics[width=0.44\textwidth]{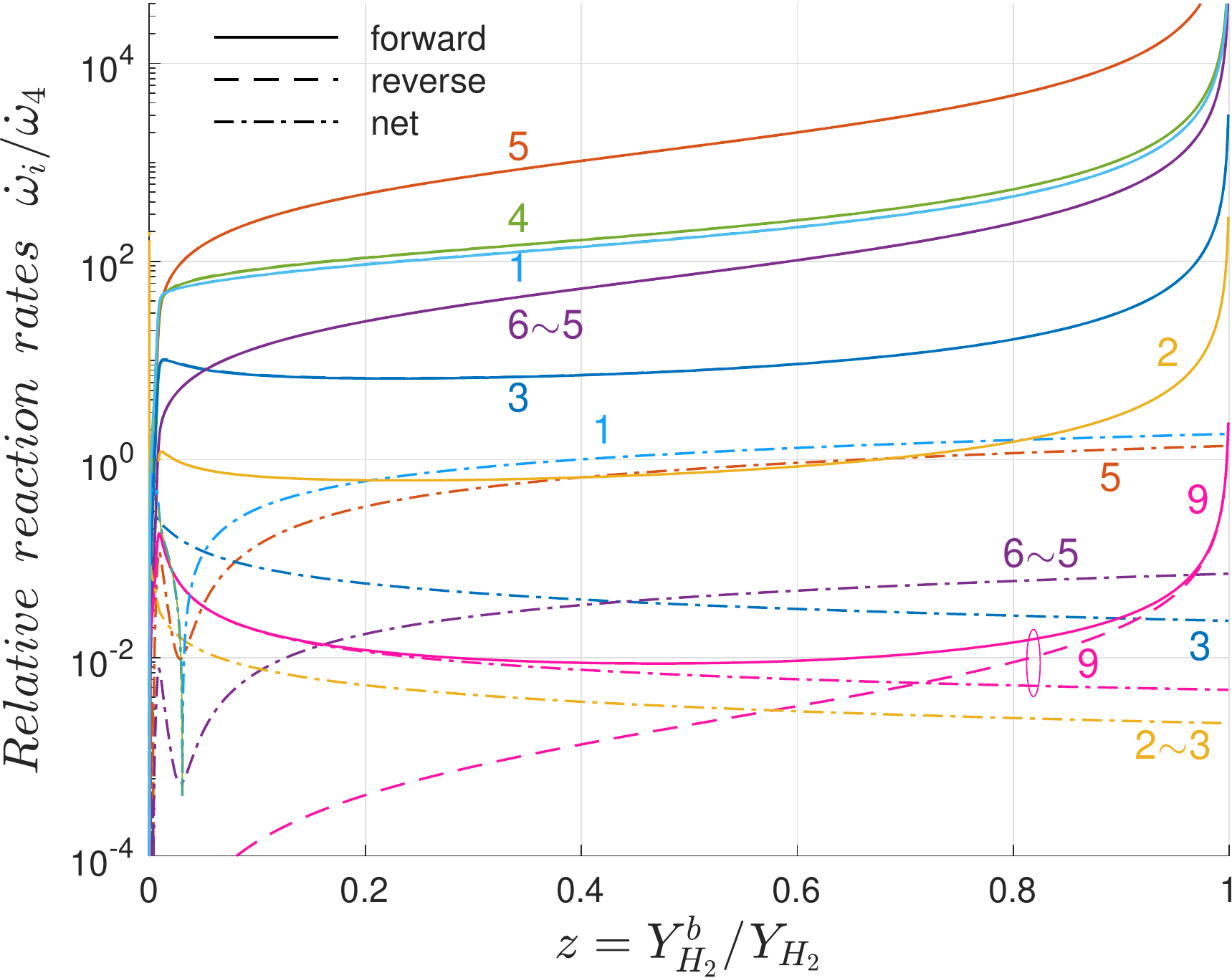} 
	\caption{Reaction rates, scaled with $\dot{\omega}_{4}$, of the most relevant reactions in the \ce{H2}-\ce{O2} mechanism (see \ref{FulH2O2Mechanism}). Solid and dash lines (coincident due to the almost exact partial equilibrium of all but reaction 9) represent forward and reverse reaction rates respectively. Dash-dot lines represent the corresponding net rates. The symbol $\sim$ links reactions with proportional reaction rates; these are reactions with the same chemical formula, but different temperature dependence rate, which therefore become proportional in the recombination region where the temperature is constant.}
	\label{ShuffleRatesPlot}
	\end{center}
\end{figure}

$\tilde{\mathscr{S}}^e_{i}$ comprises mainly the \ce{HO2} reactions, so it scales with $\dot{\omega}_{13}$. For instance, in the approximation of the reduced mechanism of Table \ref{KineticMechanism}, $\tilde{\mathscr{S}}^e_{\ce{O}} = -\tilde{\omega}_{17}$, $\tilde{\mathscr{S}}^e_{\ce{OH}} = \tilde{\omega}_{19} -\tilde{\omega}_{17}$ and $\tilde{\mathscr{S}}^e_{\ce{H}} = -\tilde{\omega}_{13}$. However, other out-of-equilibrium reactions not involving \ce{HO2}, such as reaction 9, H + O + M $\lra$ OH + M, (see Figure \ref{ShuffleRatesPlot}) can be included in $\tilde{\mathscr{S}}^e_{i}$.

At variance, the shuffle reactions rates $\tilde{\omega}_{1,4,5}$ are left out of $\tilde{\mathscr{S}}^e_{i}$ because they are in almost exact partial equilibrium, with $\tilde{\omega}_{1,4,5} \approx 0$ -- used to derive \eqref{PartialEquilibriumSolvedOnlyH2} --, and need to be evaluated in the next order of approximation.

To do so, the scaled mass fractions $\psi_i = Y_{i}/Y^e_{i}$ of $i = $ \radicals are introduced as independent variables instead of $Y_i$. $\tilde{\omega}_{1,4,5}$ can then be written, as shown in Appendix C
, as algebraic functions of $\psi_i$
\begin{subequations}
	\label{ShuffleRates}
	\begin{gather}
		\tilde{\omega}_{1} = \tilde{\Omega}^e_{1} \left( \psi_{\ce{H}} - \psi_{\ce{O}} \psi_{\ce{OH}}\right), \\
		\tilde{\omega}_{4} = \tilde{\Omega}^e_4 \left(\psi_{\ce{OH}} - \psi_{\ce{H}} \right), \\
		\tilde{\omega}_{5} = \tilde{\Omega}^e_5 \left( \psi_{\ce{OH}}^2 - \psi_{\ce{O}} \right) ,
	\end{gather}
\end{subequations}
with the coefficients $\tilde{\Omega}^e_i = \dot{\omega}^{e}_{i,f}/\dot{\omega}^{e}_{13,f}$, functions of only $z$.

It follows from these expressions and from $\tilde{\Omega}^e_i \gg 1$ (see \ref{ShuffleRatesApp}) that the new variables $\psi_i$ are such that $|\psi_i - 1| \ll 1$, which confirms that the equilibrium concentrations \eqref{PartialEquilibriumSolvedOnlyH2} are excellent approximations. They can be used in particular to evaluate the derivatives on the left hand sides of \eqref{MainEquationsDimensionless} as
\begin{gather}
	\partial X_{k} / \partial \xi \approx \acute{X}^e_{k} \cdot \partial X_{\ce{H2}} / \partial \xi,
	\label{DerivativestoXH2}
\end{gather}
with $\acute{X}^e_{k} = \partial X^e_{k}/\partial X_{\ce{H2}}$ the derivatives with respect to $X_{\ce{H2}}$, functions in the first approximation of only $z$.

The system \eqref{MainEquationsDimensionless} can thus be written as
\begin{subequations}
	\label{MainSystemFinal}
	\begin{gather}
		\label{MainSystemFinalH2}
		\frac{\partial \ln Y_{\ce{H2}}} {\partial \xi}  = - \tilde{\omega}_{4} , \\
		\label{MainSystemFinalO}
		- \tilde{\omega}_{4} \acute{X}^e_{\ce{O}} 
			= \tilde{\omega}_{5} + \tilde{\omega}_{1} + \tilde{\mathscr{S}}^e_{\ce{O}}, \\
		\label{MainSystemFinalOH}
		- \tilde{\omega}_{4}\acute{X}^e_{\ce{OH}} = \tilde{\omega}_{1} - \tilde{\omega}_{4} - 2 \tilde{\omega}_{5} + \tilde{\mathscr{S}}^e_{\ce{OH}}, \\
		\label{MainSystemFinalH}
		- \tilde{\omega}_{4}\acute{X}^e_{\ce{H}} = \tilde{\omega}_{4} - \tilde{\omega}_{1} + \tilde{\mathscr{S}}^e_{\ce{H}}, \\
		\label{MainSystemFinalHO2}
		\tilde{\mathscr{S}}^e_{\ce{HO2}} = 0,
	\end{gather}
\end{subequations}
where the equations for \radicals have been divided by \eqref{H2ME} to eliminate $\partial X_{\ce{H2}} / \partial \xi$. Additionally, the last equation \eqref{MainSystemFinalHO2} is written taking into account that \ce{HO2} is in steady state, as shown in \ref{HO2RatesApp}, so $\partial X_{\ce{HO2}}/\partial \xi \approx 0$.

This formulation \eqref{MainSystemFinal}, in terms of $Y_{\ce{H2}}$ for \ce{H2} and of $\psi_i$ for the radicals \radicals as dependent variables, is a differential-algebraic system involving explicitly the spatial derivative of $Y_{\ce{H2}}$. In contrast, $\psi_i$ enter only implicitly through the rates $\tilde{\omega}_{1, 4, 5}$. These rates can thus be used temporarily as independent variables in place of $\psi_i$. Once the rates $\tilde{\omega}_{1, 4, 5}$ are determined, as done next, $\psi_i$ follow from \eqref{ShuffleRates}.

Thus, the equations for \radicals represent three linear algebraic relations from which the rates $\tilde{\omega}_{1, 4, 5}$ can be solved explicitly in terms of $\tilde{\mathscr{S}}^e_{i}$ and $\acute{X}^e_{i}$ as done in \ref{ShuffleRatesApp}, giving in particular
\begin{gather}
	\label{Omega4Solved}
	\tilde{\omega}_{4} \approx
		- \frac{  \tilde{\mathscr{S}}^e_{\ce{OH}} + 2\tilde{\mathscr{S}}^e_{\ce{O}} + 3\tilde{\mathscr{S}}^e_{\ce{H}} }
		{ \acute{X}^{e}_{\ce{OH}} + 2(\acute{X}^{e}_{\ce{O}} + 1) + 3\acute{X}^{e}_{\ce{H}} }.
\end{gather}

This expression for $\tilde{\omega}_{1}$ and  also $\tilde{\omega}_{4,5}$ are proportional to $\tilde{\mathscr{S}}^e_{\ce{i}}$, and therefore to the \ce{HO2} reactions rates, i.e. $\tilde{\omega}_{1,4,5} \propto \tilde{\omega}^e_{13}$, thus proving the statement, anticipated above without proof, that the \ce{HO2} kinetics governs the deviations from exact partial equilibrium of the shuffle reactions. 

The approximation \eqref{Omega4Solved} is general regarding the \ce{HO2} kinetics used to evaluate $\tilde{\mathscr{S}}^e_{i}$. Since all these rates depend ultimately on the concentrations of \radicals, which are obtained from \eqref{PartialEquilibriumSolvedOnlyH2}, including more reactions, or even of the \ce{H2O2} kinetics, is thus straightforward save for the extended required calculations.

\begin{figure}[h!]
	\begin{center}
		\includegraphics[width=0.44\textwidth]{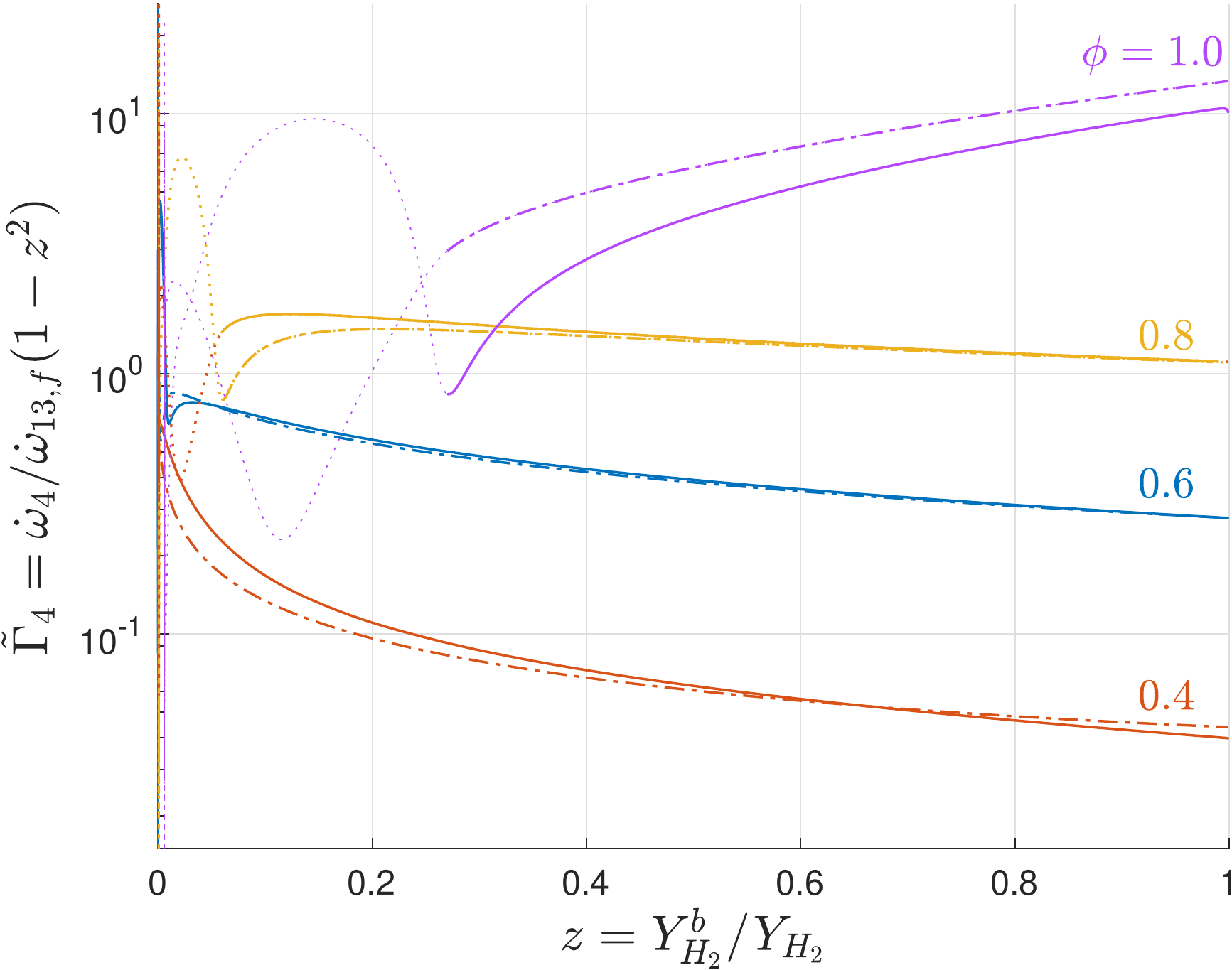} 
		\caption{Distribution of the normalized $\dot{\omega}_4$ for several representative values of $\phi$, showing that $\dot{\omega}_4$ scales with $\dot{\omega}_{13} \sim \dot{\omega}_{13, f}(1-z^2)$, with the factor $1-z^2$ accounting for the zero net rate at $z = 1$; $\dot{\omega}_{13, f}$ carries the order of magnitude of the rate and accounts for  the effect of $\phi$. The dash-dot lines represent the approximation \eqref{Omega4Solved} evaluated with the full (exact) chemistry, showing that it is excellent around $\phi = 0.6$. The dotted lines represent, for reference, the fuel consumption layer taken to roughly occur for $x < 0.5$ mm.}
		\label{Rate4}
	\end{center}
\end{figure}

Using $\tilde{\omega}_{4}$ back in \eqref{MainSystemFinalH2} leads finally to the differential equation governing the distribution of \ce{H2}
\begin{gather}
	\label{H2FinalEqDimensionless}
	\frac{\partial \ln z} {\partial \xi} 
		= \frac{  \tilde{\mathscr{S}}^e_{\ce{OH}} + 2\tilde{\mathscr{S}}^e_{\ce{O}} + 3\tilde{\mathscr{S}}^e_{\ce{H}} }
		 { \acute{X}^{e}_{\ce{OH}} + 2(\acute{X}^{e}_{\ce{O}} + 1) + 3\acute{X}^{e}_{\ce{H}} }.
\end{gather}
or in terms of the physical variables
\begin{gather}
	\label{H2FinalEqPhysical}
	\frac{\partial \ln Y_{\ce{H2}}} {\partial x}  =  
		- \frac{ \mathcal{M}_{\ce{H2}} } { \dot{m}_L }
		\frac{ \dot{\mathscr{S}}^e_{\ce{OH}} + 2\dot{\mathscr{S}}^e_{\ce{O}} + 3\dot{\mathscr{S}}^e_{\ce{H}} }
		 { \acute{X}^{e}_{\ce{OH}} + 2(\acute{X}^{e}_{\ce{O}} + 1) + 3\acute{X}^{e}_{\ce{H}} }
		,
\end{gather}

The source term of \eqref{H2FinalEqDimensionless}, function of only $z$ (of $Y_{\ce{H2}}$ in \eqref{H2FinalEqPhysical}) in the first approximation -- so these are simply quadratures --, has the form (see \ref{HO2RatesApp})
\begin{gather}
	\label{Gamma4Source}
	\tilde{\omega}_{4} = \dot{\omega}_{4}/\dot{\omega}_{13,f} = \tilde{\Gamma}_4(z)(1-z^2),
\end{gather}
with $\tilde{\Gamma}_4(z)$ a regular, finite function of order unity, plotted for representative values of $\phi$ in Figure \ref{Rate4}.

Equation \eqref{H2FinalEqDimensionless} together with \eqref{Gamma4Source} give the asymptotic form of the decay to equilibrium as $1 - z \propto e^{-\xi/\Lambda}$ for $\xi \rightarrow \infty$, with a dimensionless length scale $\Lambda = (2\tilde{\Gamma}_4^b)^{-1}$, monotonously decreasing with $\phi$ as follows from the behavior of $\tilde{\Gamma}_4^b = \tilde{\Gamma}_4(z=1)$ displayed in Figure \ref{Rate4}. The $\phi$-dependence of $\Lambda$ thus confirms the larger thickness of the recombination region of leaner flames as anticipated in Figure \ref{LengthsAndPeclets}.

The equation \eqref{H2FinalEqDimensionless} (and \eqref{H2FinalEqPhysical}) requires an initial condition giving the starting value of \ce{H2}. This initial condition is the matching condition with the solution upstream of the recombination region and is obtained as the constant, in the first approximation, value of \ce{H2} amount left downstream of the fuel consumption layer, after the shuffle reactions reach partial equilibrium, with the rates of the \ce{HO2} reactions set to zero to prevent its slow decay. Instead, to explore the errors associated with the approximations discussed previously and to remove as much as possible errors associated with matching, the initial condition is taken as the value of the exact solution at some specific location downstream of the fuel consumption layer. As shown below, the errors associated with the approximate solution decrease as the initial condition is chosen at larger distances downstream of the fuel consumption layer. This is so because the difference between the exact and the approximate value of the slope $\partial \ln z/\partial \xi$ decreases approaching the burnt equilibrium where the difference is zero. Thus, an upper bound of the errors of the simplified model can be obtained by choosing the initial close to the fuel consumption layer as done below.

%
\begin{figure}[h!]
	\begin{center}
		\includegraphics[width=0.44\textwidth]{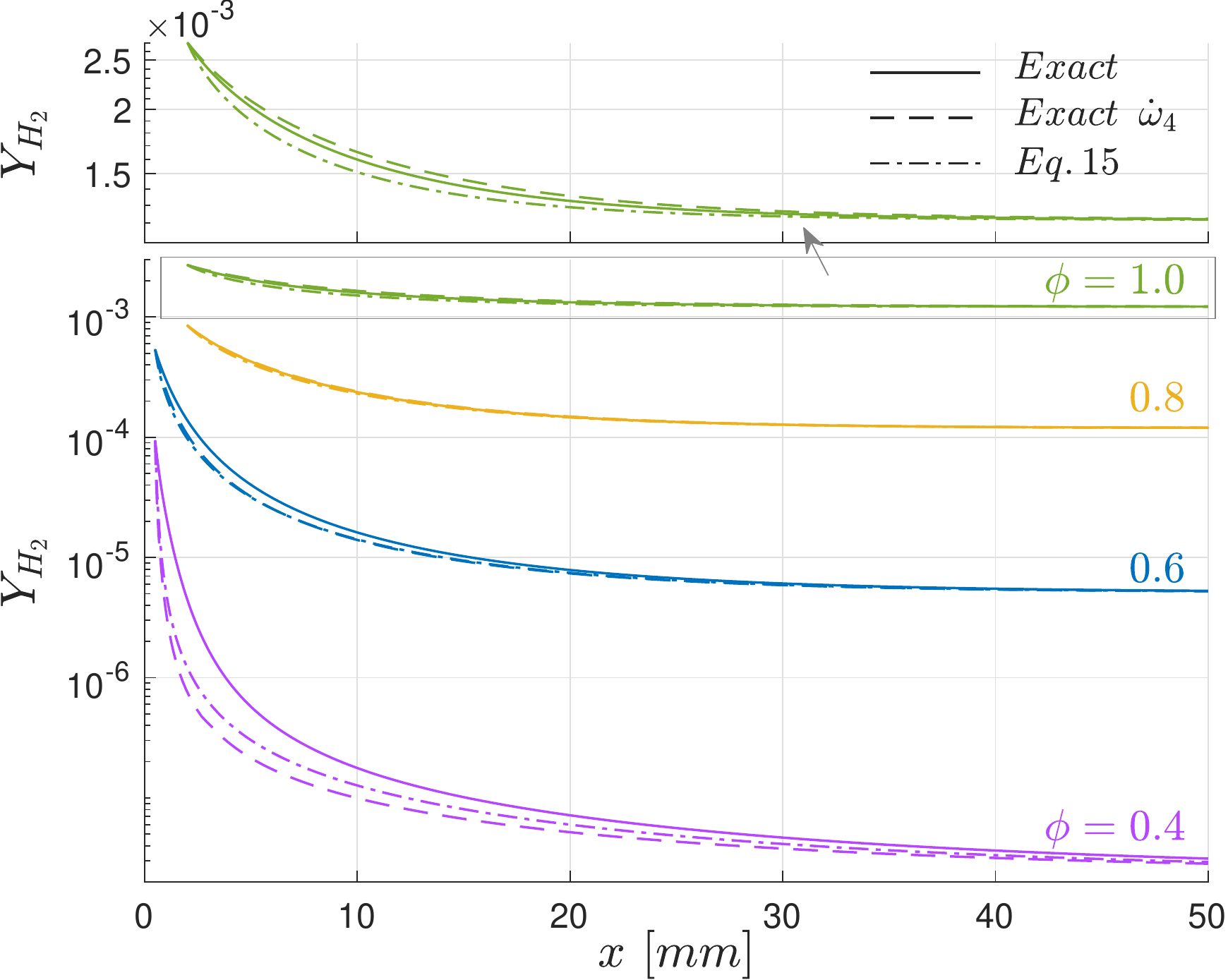} \hspace{1cm}
		\includegraphics[width=0.44\textwidth]{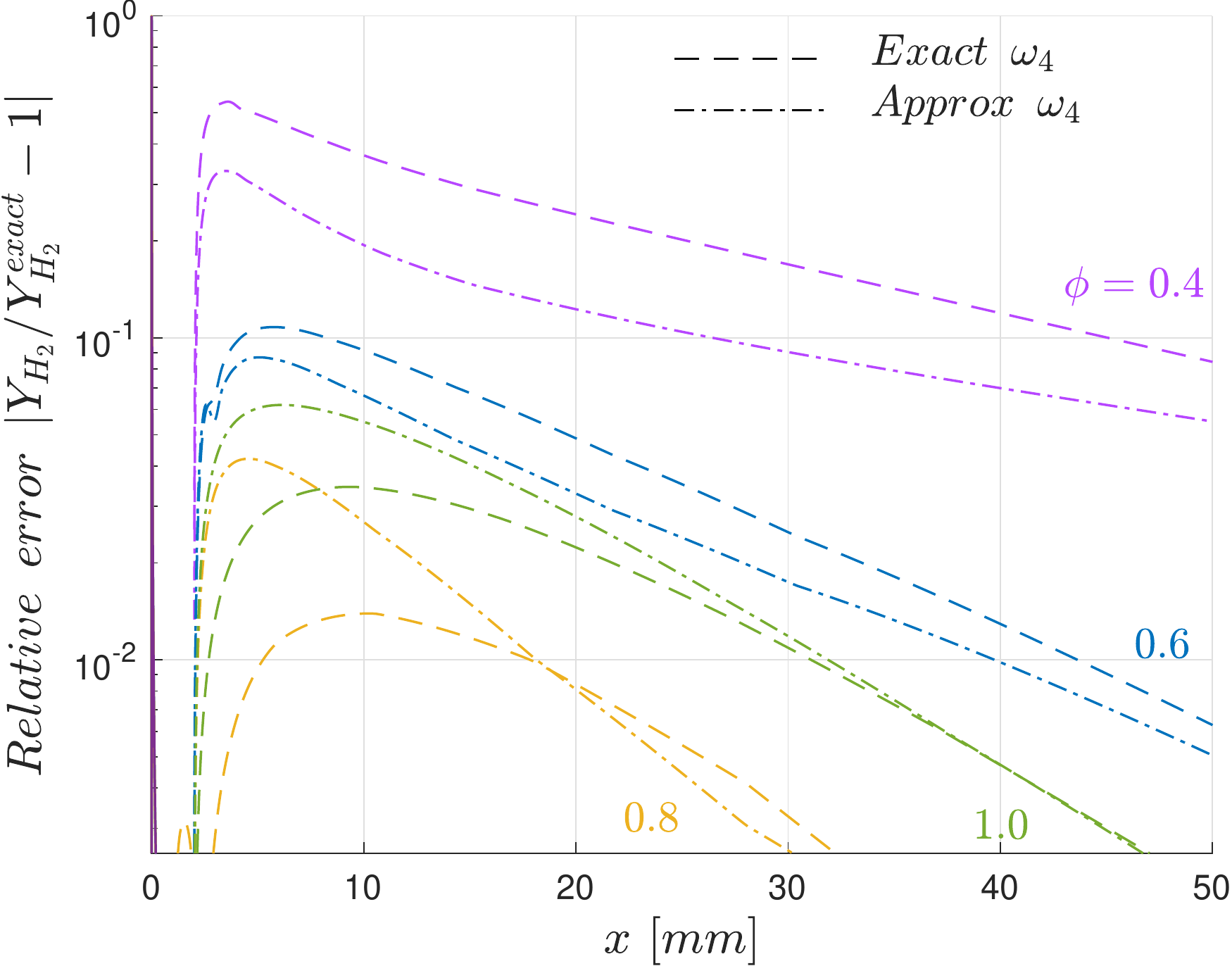} 
		\caption{The left plot compares the numerical solution of the full problem, labelled as {\it{Exact}} (solid lines), with the solution of \eqref{MainSystemFinalH2} using i) the exact rate $\dot{\omega}_4$ (dash lines) and ii) evaluating exactly the approximation \eqref{Omega4Solved} of $\dot{\omega}_4$ (dash-dot lines). The right plot represents the corresponding relative errors. The initial condition to integrate \eqref{H2FinalEqDimensionless} has been (arbitrarily) chosen as the value of $Y_{\ce{H2}}(x_0)$ from the exact solution, with $x_0$ = 0.5 mm, which is roughly where the recombination region starts. As discussed in Figure \ref{Solutionphi06}, choosing to start the solution further downstream decreases the relative error.}
		\label{ConvectionReduced}
	\end{center}
\end{figure}

\subsection{Results.}

In order to illustrate the validity of \eqref{MainSystemFinalH2} and of \eqref{H2FinalEqDimensionless}, Figure \ref{ConvectionReduced} compares the exact -- numerical solution of the full flame problem -- distribution of $Y_{\ce{H2}}$, with solutions of \eqref{MainSystemFinalH2} and \eqref{H2FinalEqDimensionless} integrated using the exact numerical solution to evaluate their right hand sides. The corresponding relative errors measure in the first case the effect of neglecting the diffusive transport in the full problem; whereas in the second case the errors also include the additional effect of approximating $\tilde{\omega}_{4}$ with \eqref{Omega4Solved}.

\begin{figure}[h!]
	\begin{center}
		\includegraphics[width=0.44\textwidth]{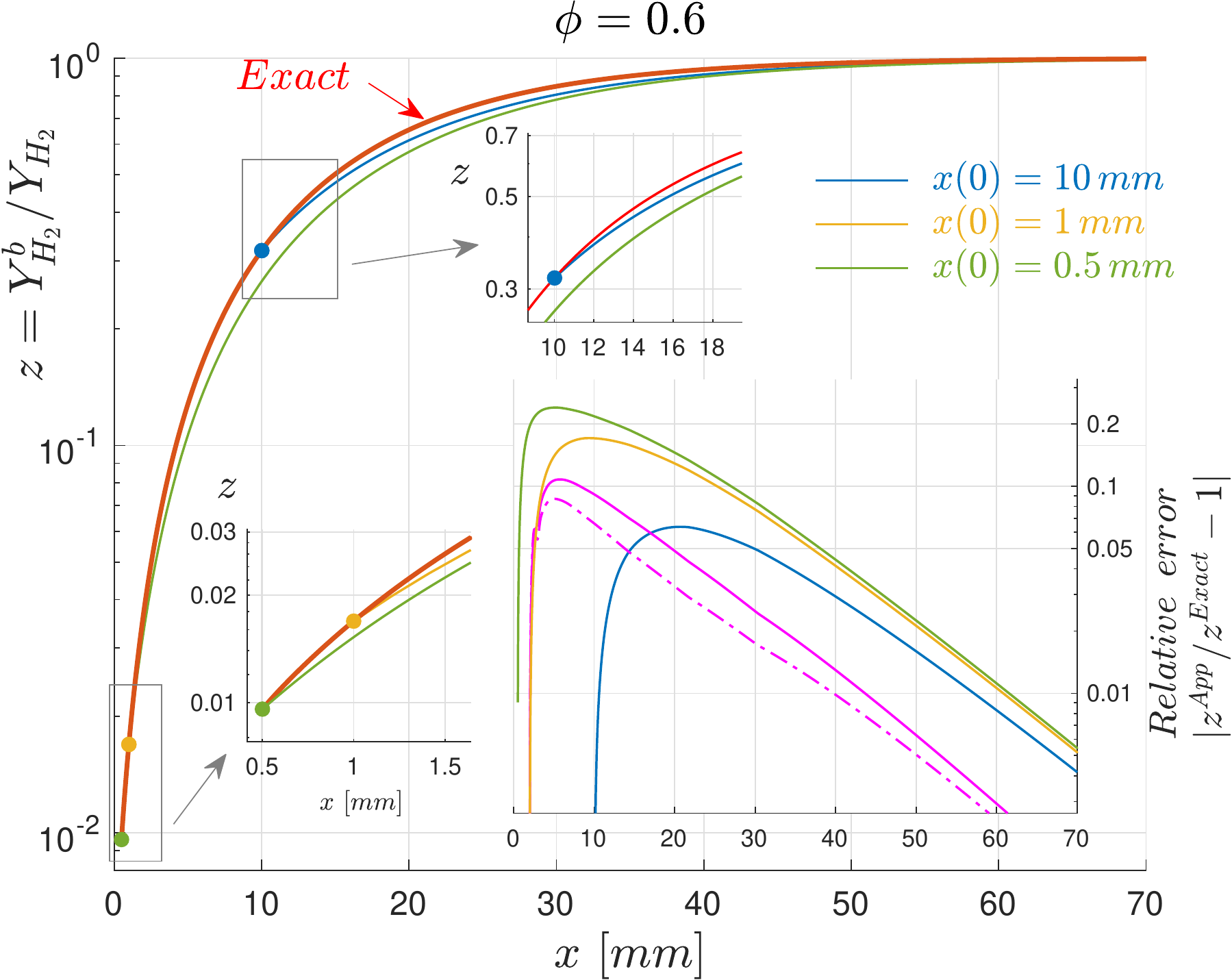} \hspace{1cm}
		\includegraphics[width=0.44\textwidth]{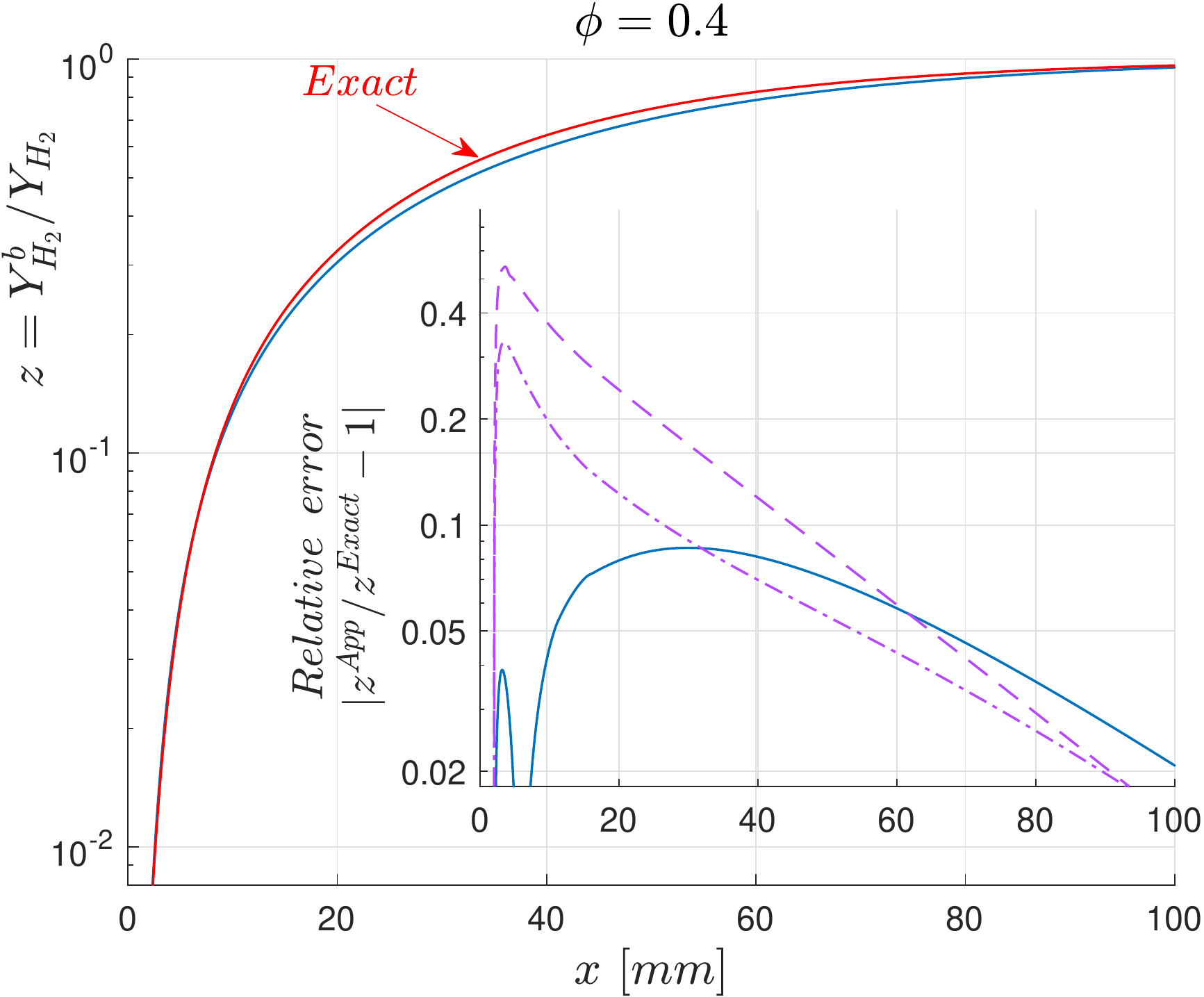}
		\caption{On the left, the solution, for $\phi = 0.6$, of \eqref{H2FinalEqDimensionless} using the reduced \ce{HO2} mechanism of Table \ref{KineticMechanism} and the expressions for the rates, based on \eqref{PartialEquilibriumSolvedOnlyH2}, given in \ref{HO2RatesApp}. To understand the effect of the initial conditions $Y^0_{H_2}(x_0)$, the solutions for  $x_0 = 0.5$ mm (green line), $x_0 = 1$ mm (yellow line) and $x_0 = 10$ mm (blue), with $Y^0_{H_2}$ evaluated from the exact solution, are included. The inset on the lower right corner represents the associated errors; for reference, also shown (purple lines) the errors for $\phi = 0.6$ from Figure \ref{ConvectionReduced}. On the right, the solution and corresponding error for $\phi = 0.4$ using the same mechanism and $x_0 = 1.0$.}
		\label{Solutionphi06}
	\end{center}
\end{figure}

These results show that neglecting the diffusive transport is a good approximation in not too lean flames but becomes inaccurate, as Figure \ref{LengthsAndPeclets} anticipated, especially early in the recombination region, in slower flames. 
On the contrary, the approximation \eqref{Omega4Solved}, based on constant values of the temperature and \ce{O2} concentration provides an excellent approximation not too close to stoichiometry, as expected from Figure \ref{H2RadicalsSpecies}. 

On the other hand, Figure \ref{Solutionphi06} shows the solution of \eqref{H2FinalEqDimensionless} for $\phi = 0.4$ and 0.6 using the reduced mechanism of Table \ref{KineticMechanism}, i.e. $\tilde{\mathscr{S}}^e_{\ce{O}} = -\tilde{\omega}_{17}$, $\tilde{\mathscr{S}}^e_{\ce{OH}} = \tilde{\omega}_{19} -\tilde{\omega}_{17}$ and $\tilde{\mathscr{S}}^e_{\ce{H}} = -\tilde{\omega}_{13}$, with the rates $\tilde{\omega}_{13, 17, 19}$ evaluated as functions of $z$ using the expressions given in \ref{HO2RatesApp}. The corresponding errors, with a peak of around 20\%, include the additional effect of using the reduced kinetics instead of the full mechanism (purple lines), and also of using \eqref{PartialEquilibriumSolvedOnlyH2}.

\section{Discussion and conclusions.}\label{Conclusions}

The structure and kinetics of the recombination region has been studied in detail in lean \ce{H2} flames, with the main result that this region can be described using the \ce{H2} concentration as only degree of freedom. Its distribution is given by the solution of the first-order ordinary differential equation \eqref{H2FinalEqDimensionless}, or \eqref{H2FinalEqPhysical} in physical variables, whereas the \radicals distributions follow from that of \ce{H2} using the equilibrium relation \eqref{PartialEquilibriumSolvedOnlyH2}. 

\begin{figure}[h!]
	\begin{center}
	\includegraphics[width=0.44\textwidth]{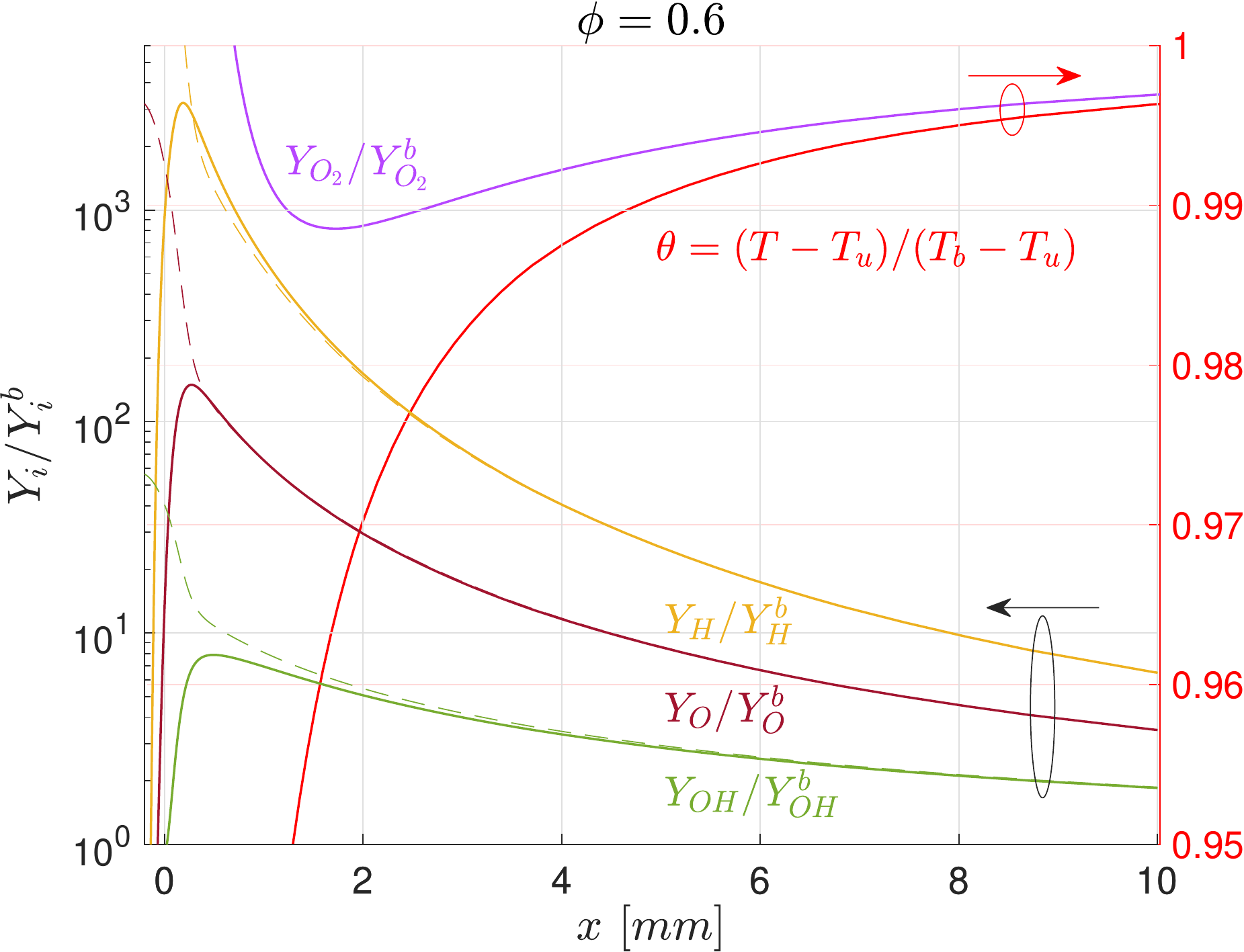} \hspace{1cm}
	\includegraphics[width=0.44\textwidth]{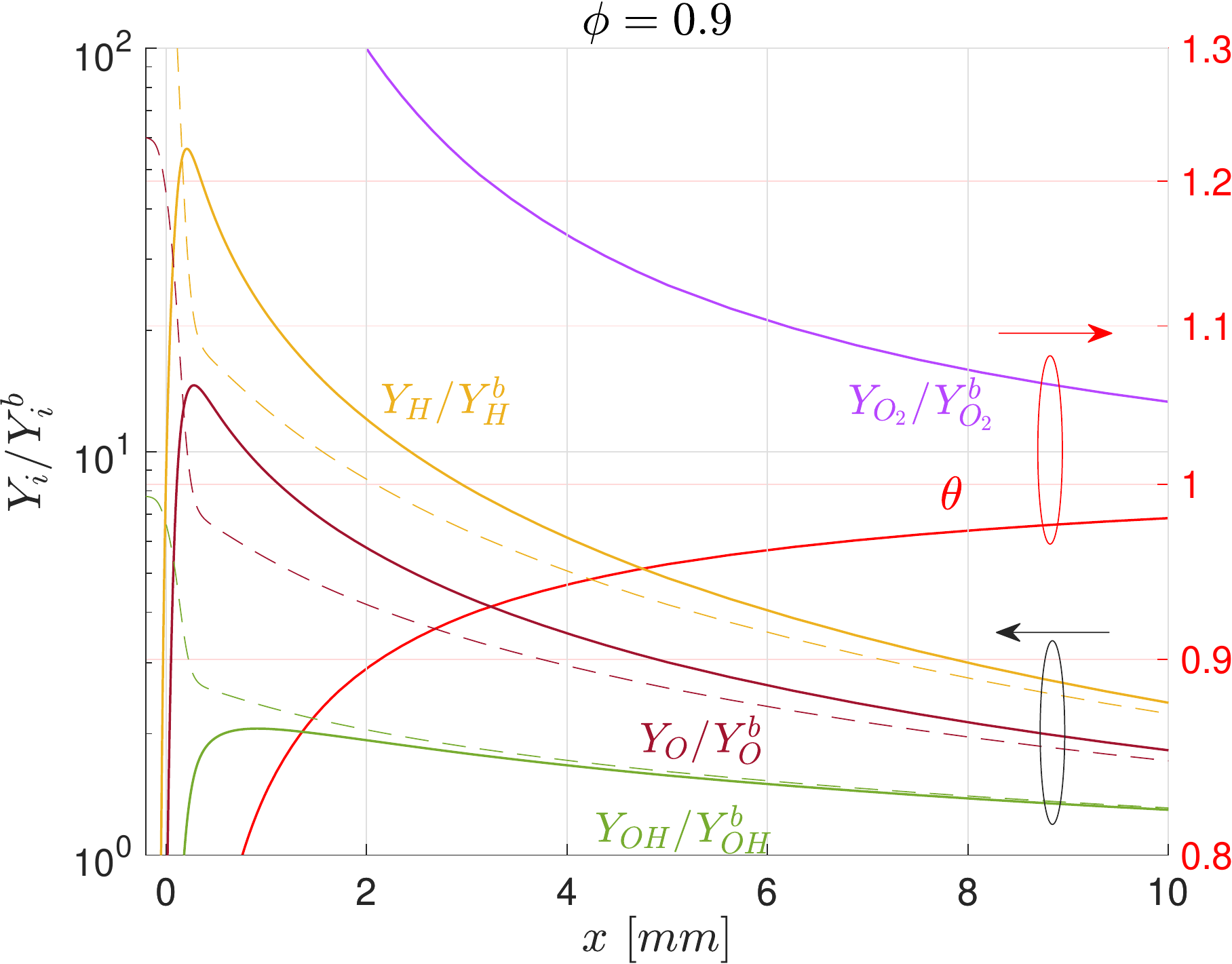} 
	\caption{Distributions of the mass fractions $Y_{i}$ of O, H, OH and \ce{O2}, normalized with the corresponding final burnt equilibrium values $Y^b_{i}$. The thin dash lines are the approximations \eqref{PartialEquilibriumSolvedOnlyH2}, based on constant values $\theta, Y_{\ce{H2O}}/Y^b_{\ce{H2O}}, Y_{\ce{O2}}/Y^b_{\ce{O2}} \approx 1$ throughout this region. These approximations are excellent in sufficiently lean flames (left plot). However, as stoichiometry is approached (right plot), the temperature and the amount of \ce{O2} significantly deviate -- notice the different vertical right scale range on each plot --, early in the recombination region, from their final equilibrium value.}
	\label{ErrorsOnlyH2}
	\end{center}
\end{figure}

\begin{figure*}[h!]
	\begin{center}
	\includegraphics[width=0.44\textwidth]{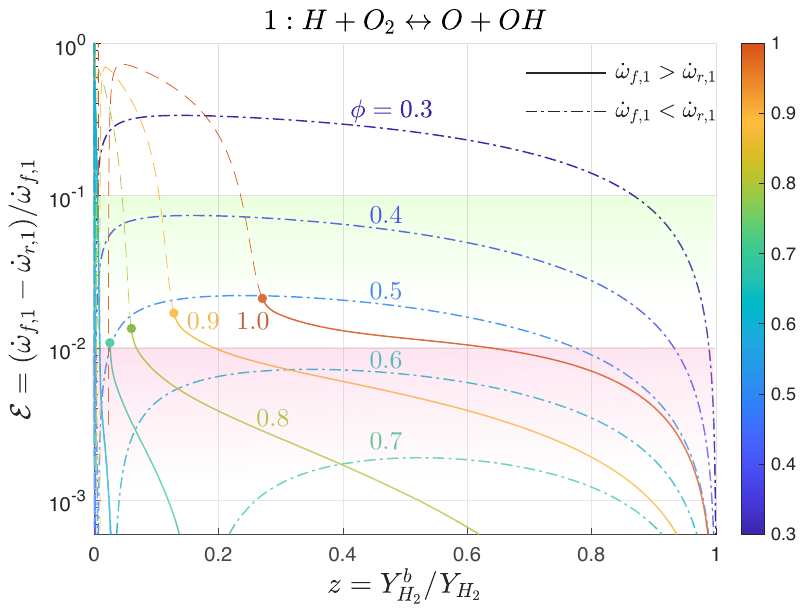} \hspace{1cm}
	\includegraphics[width=0.44\textwidth]{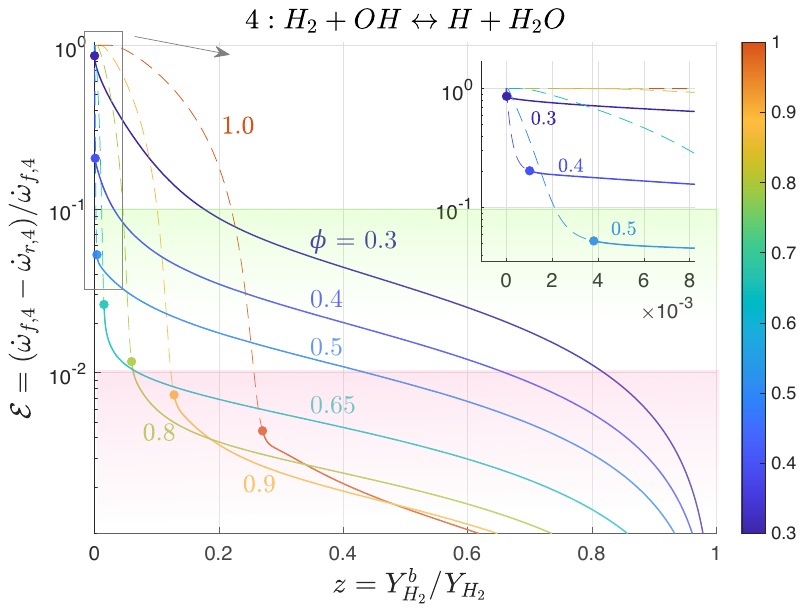} \\[.5cm]
	\includegraphics[width=0.44\textwidth]{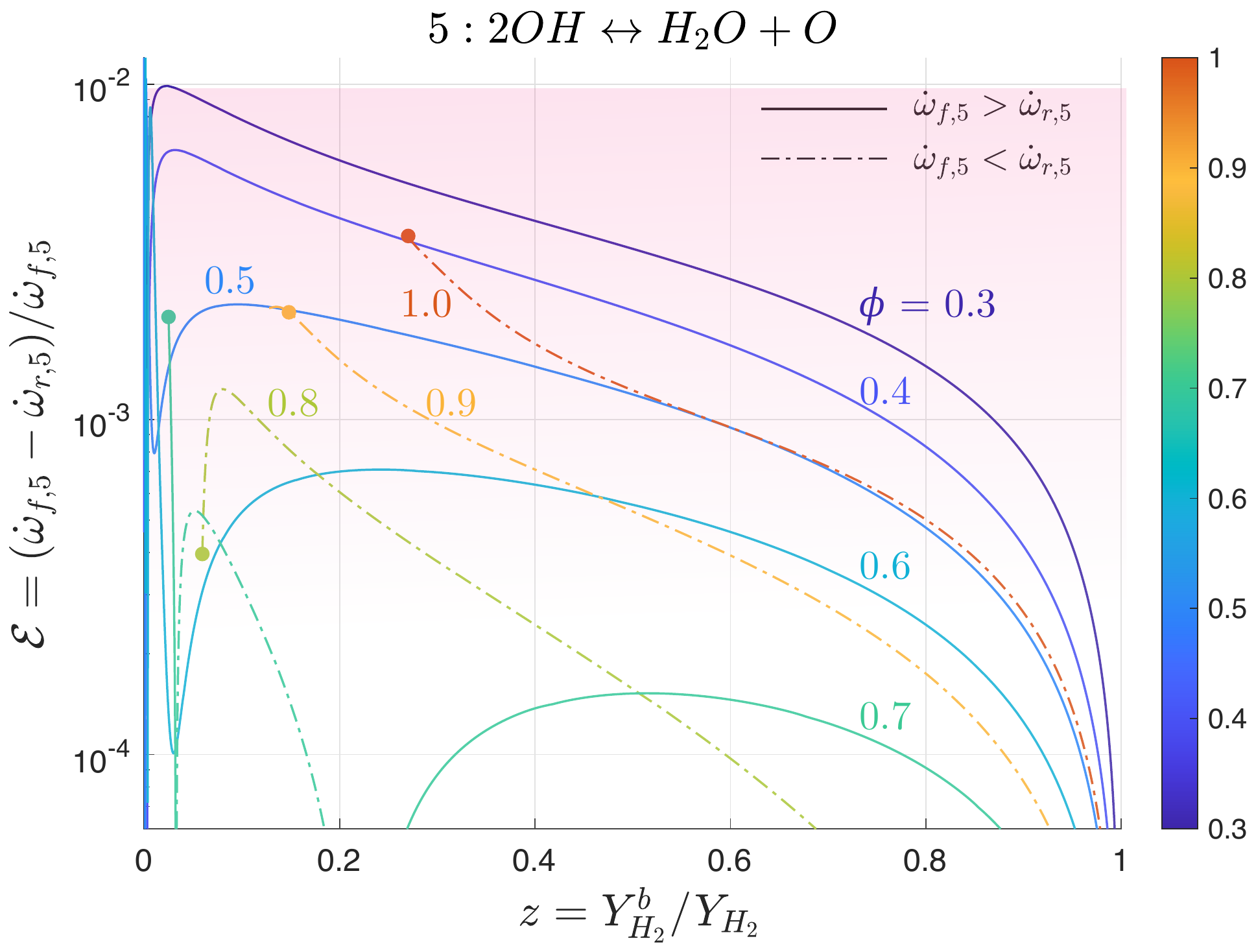}
	\caption{Deviations from exact partial equilibrium of the shuffle reactions. Solid ($\dot{\omega}_{i, f} > \dot{\omega}_{i, r}$) and dash-dot ($\dot{\omega}_{i, f} < \dot{\omega}_{i, r}$) lines represent the recombination region, whereas thin dash lines represent the region upstream of it -- fuel consumption layer and preheat region --  and are removed from the bottom plot for clarity. The green and red shadowing mark the regions with errors are smaller than 10\% and 1\% respectively. The side color bars are guides to the lines color-$\phi$ mapping.}
	\label{PartialEquilibrium}
	\end{center}
\end{figure*}

This significant simplification is based on the simultaneous partial equilibrium of the three shuffle reactions, which is accurate in not too lean flames as shown in Figures \ref{ConvectionReduced} and \ref{PartialEquilibrium}. It is based also on the assumption of constant values of the temperature and \ce{O2} and \ce{H2O} concentrations, which reach their asymptotic values quickly downstream of the fuel consumption layer in flames not too close to stoichiometry flames as illustrated in Figure \ref{ErrorsOnlyH2}. Thus, this simplified model with just one degree of freedom works well in moderately lean flames, around $\phi \approx 0.6$, but requires to include more degrees of freedom in very lean flames below $\phi < 0.4$, and close to stoichiometry. 

In the limit of very lean flames, the shuffle reaction 1, and reaction 4 to a lesser degree, cease to be in partial equilibrium as illustrated in Figure \ref{PartialEquilibrium}. The partial equilibrium of reaction 5 however still holds, furnishing a relationship between OH and H, which can be used to reduce by one the number of degrees of freedom. Additionally, H tends to be in steady state in the recombination region in these ultra lean flames 
hence eliminating the corresponding transport equation in favor of an explicit algebraic relationship providing H in terms of the rest of variables. Thus, the recombination region requires in this limit the concentrations of \ce{H2} and of O as degrees of freedom. On the other hand, despite the declining reactivity in this limit, diffusive transport needs to be accounted for, as Figures \ref{LengthsAndPeclets} and \ref{ConvectionReduced} suggest, clearly as a result of the also decreasing flame consumption rate of fresh mixture $\dot{m}''_L$, and therefore of the convective transport rate.

Close to stoichiometry, on the other hand, the transport equations for the temperature and for the \ce{O2} must be solved together with that for \ce{H2}. This is so because a non-negligible fraction of the total heat release occurs in the recombination region, so the temperature can not be assumed to be constant any more. In addition, \ce{O2} is found in these flames in amounts comparable to those of \ce{H2} and can not be assumed to be constant as seen in Figure \ref{ErrorsOnlyH2}.


\subsection{Heat losses}

The long length scale associated with the recombination region suggests that heat losses may become significant and change the structure of the recombination region (if this is not disrupted, for instance by a turbulent flow). The paper's results, however, still apply unchanged in the presence of moderate heat losses, when the induced temperature variations $\partial T_{HL}$ are small compared with the local temperature in the recombination region $\partial T_{HL}/T \approx \partial T_{HL}/T^b \ll 1$ (in similarity with, as shown in \ref{TemperatureAndProducts}, the small temperature changes induced by the combustion of the unburned \ce{H2} left past the fuel consumption layer). For instance, \ce{H2O} in the products induce heat radiation losses of the order of $\partial T/\partial x \sim X_{\ce{H2O}}\kappa_{\ce{H2O}} \sigma T^4/(\dot{m}''_L c_p) \sim 10^2$K/m, with $\kappa_{\ce{H2O}} \sim 1m^{-1}$ the \ce{H2O} mean absorption coefficient\cite{HITRAN, COELHO2018105} and $\sigma$ the Stefan-Boltzmann constant. This heat losses rate induce in turn, over the thickness of the recombination region (see Figure \ref{LengthsAndPeclets}), temperature variations such that $\partial T_{HL}/T \sim 10^{-2}$, which can therefore be neglected.

However, more intense heat losses can change the structure. For instance, Joulin and Clavin \cite{JOULIN1979139} showed that unstretched flames are led to extinction when the heat loses are increased to produce gradients on the products side of order $\partial T/\partial x \sim \beta^{-1} (T^b - T^u)/\delta_L$, with $\delta_L$ the flame thickness (without the recombination region) and $\beta \sim 10$, the Zeldovich number\footnote{$\beta = E(T^b - T^u)/R{T^b}^2$ is the Zeldovich number, which is not well defined in non-Arrhenius kinetics; it can however be estimated as $\beta \sim d \ln U_L/d \ln T^b \sim 10$ not too close to the flammability limits (where it diverges).}. Heat losses of this order induce, over the thickness of the recombination region, temperature variations such that $\partial T_{HL}/T \sim \beta^{-1} (T^b - T^u)/T^b (\mathcal{L}/\delta_L)$, which is of order unity, or even larger in slow flames. 

Thus, before extinction, heat loses such that $\partial T_{HL}/T \sim 1$ can occur, inducing changes in the recombination region structure. In this case however, the equation for the temperature would be decoupled from the species distributions (ignoring the small effect on minor species on properties such as the mixture-averaged heat conductivity or specific heat). The problem for the \ce{H2}, \radicals distributions would not be autonomous anymore because the imposed temperature distribution, function of the coordinate along the recombination region, would enter for instance in the equilibrium constants of the shuffle reactions in the conditions \eqref{PartialEquilibriumSolved}. In addition, low temperatures could bring the shuffle out of equilibrium, inducing effects similar to those previously discussed for very lean flames. Thus, even in this case of strong heat losses, when $\partial T_{HL}/T \sim 1$, these can be accounted for with modifications similar to those considered above to extend the validity to extreme values of $\phi$, so the main results still apply.

\section{Appendixes.}\label{Appendixes}

\appendix

\section{Temperature and products.} \label{TemperatureAndProducts}

The equations for $\theta$, \ce{H2O} and \ce{O2}, are
\begin{subequations}
	\label{ProductsDimensionless}
	\begin{gather}
		\label{TempEq}
		\frac{ 1 } {X_{\ce{H2}}}\frac{\partial \theta} {\partial \xi} = \sum_i{q_i \tilde{\omega}_i } , \\ 
		\label{O2ME}
		\frac { 1 } {X_{\ce{H2}}} \frac{\partial X_{\ce{O2}}} {\partial \xi} = \tilde{\omega}_{1} + \tilde{\mathscr{S}}^e_{\ce{O2}}, \\
		\label{H2OME}
		\frac { 1 } {X_{\ce{H2}} } \frac{\partial X_{\ce{H2O}}} {\partial \xi} = \tilde{\omega}_{4}  + \tilde{\omega}_{5} + \tilde{\mathscr{S}}^e_{\ce{H2O}},
	\end{gather}
\end{subequations}
with $\theta = (T-T^u)/(T^b-T^u)$, the temperature increment over the fresh mixture temperature $T^u$, and $q_i$ the molar heat release from each reaction $i$, scaled with $\mathcal{M}_m c_{p, m}(T^b-T^u)$ with $\mathcal{M}_m$ and $c_{p, m}$ the mixture molar mass and heat capacity per unit mass respectively.


Both sides of these equations are of order unity and therefore the variations of these variables are proportional to the local amount $X_{\ce{H2}}$ of \ce{H2} being burned, $\partial ( \theta, X_{\ce{O2}}, X_{\ce{H2O}}) \sim X_{\ce{H2}} \ll \theta, X_{\ce{O2}}, X_{\ce{H2O}}$, which holds not too to stoichiometry as shown for instance in Figure \ref{H2RadicalsSpecies}. 



\section{\ce{HO2} reaction rates.}\label{HO2RatesApp}

The \ce{HO2} reactions are out of equilibrium in the recombination region and can therefore be evaluated in the first approximation, using the equilibrium concentrations $\C{O}^{e}$, $\C{OH}^{e}$, $\C{H}^{e}$ from \eqref{PartialEquilibriumSolvedOnlyH2}. The resulting approximate rates carry errors due to the non-zero deviations from partial equilibrium of the shuffle reactions (see Figure \ref{PartialEquilibrium}) and to the assumption of constant values of T, \C{O2} and \C{H2O}.


The forward rate of reaction 13, the rates scale, can be written, with $\dot{\omega}^b_{13, f} = (k_{13,f}\C{O2}\C{H})|^{b}$, as 
\begin{gather}
	\label{HO2Rates13}
	\dot{\omega}_{13, f} = k_{13,f}\C{O2}\C{H} \approx \dot{\omega}^b_{13, f}/z^{3/2}.
\end{gather}
%

\begin{figure}[h!]
	\begin{center}
	\includegraphics[width=0.44\textwidth]{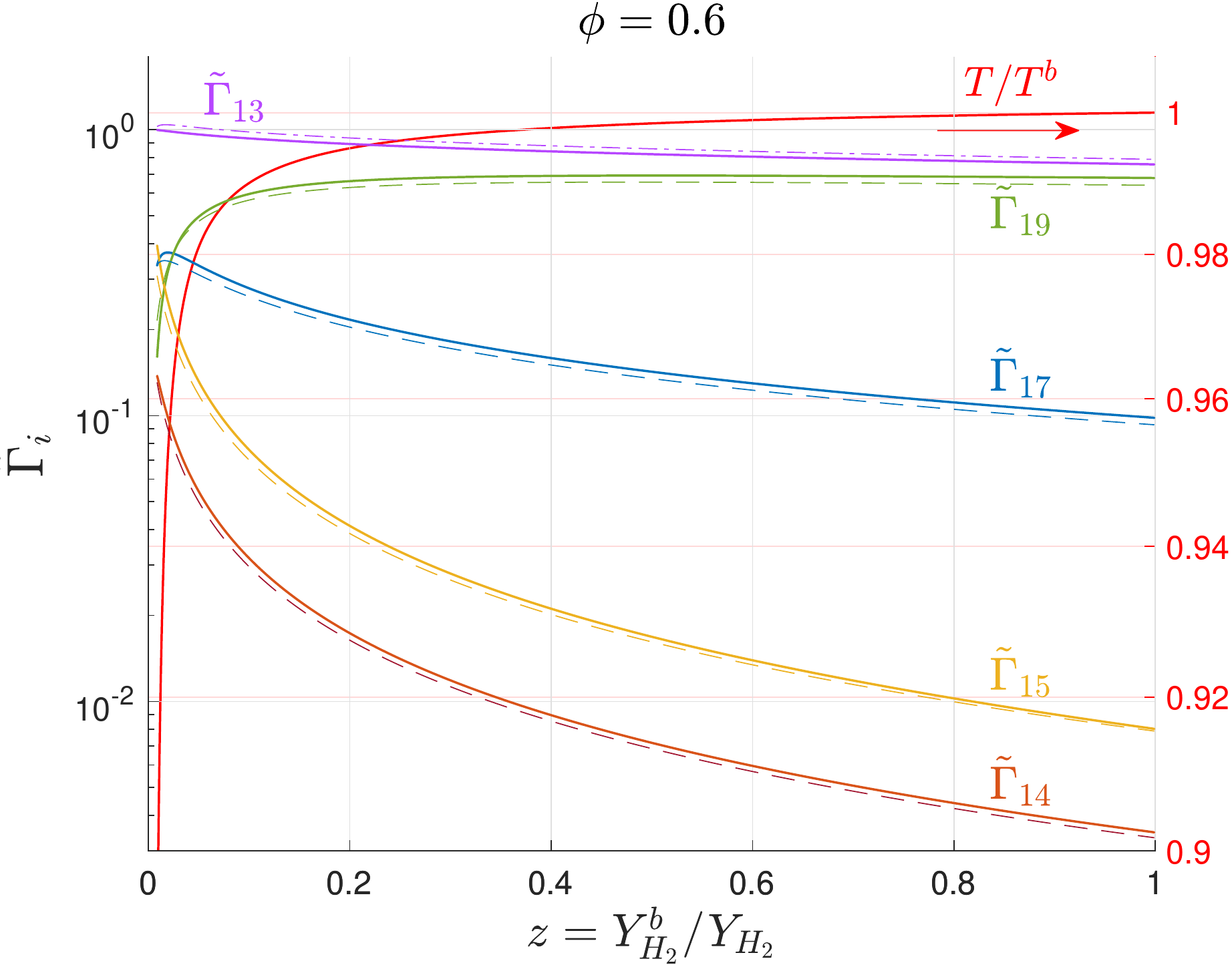} 
	\caption{Distributions of the functions $\tilde{\Gamma}_m$  \eqref{HO2Ratesm} for the more relevant reactions of the \ce{HO2} mechanism for $\phi = 0.6$. The dash lines are the approximations \eqref{DinHO2Rates}. $\tilde{\Gamma}_{13} =  \sum_{m\ne13}{\tilde{\Gamma}_{m}}$, which shows that the reactions 17 and 19 are dominant in most of the recombination region. Reactions 14 and 15 become relevant at the interface with the fuel consumption layer, because they are dominant inside it, as shown by Fernández-Galisteo et al. \cite{DaniHydrogen,DaniHydrogenI}, who use reactions 14, 15 and 19 (their 5, 6, 7 respectively) to model \ce{HO2} in the fuel consumption layer; note however that they consider these reactions as irreversible.}
	\label{BigGammasWith1415App}
	\end{center}
\end{figure}

On the other hand, for any \ce{HO2} reaction,
\begin{gather}
	\label{HO2Ratesm}
	\tilde{\omega}_{m} = \dot{\omega}_{m}/ \dot{\omega}_{13, f} \approx \tilde{\Gamma}_m(z) (1 - z^2),
\end{gather}
with $\tilde{\Gamma}_m(z)$ finite, well-behaved,  order unity functions of $z$ plotted in Figure \ref{BigGammasWith1415App} for representative m.

\begin{figure}[h!]
	\begin{center}
		\includegraphics[width=0.44\textwidth]{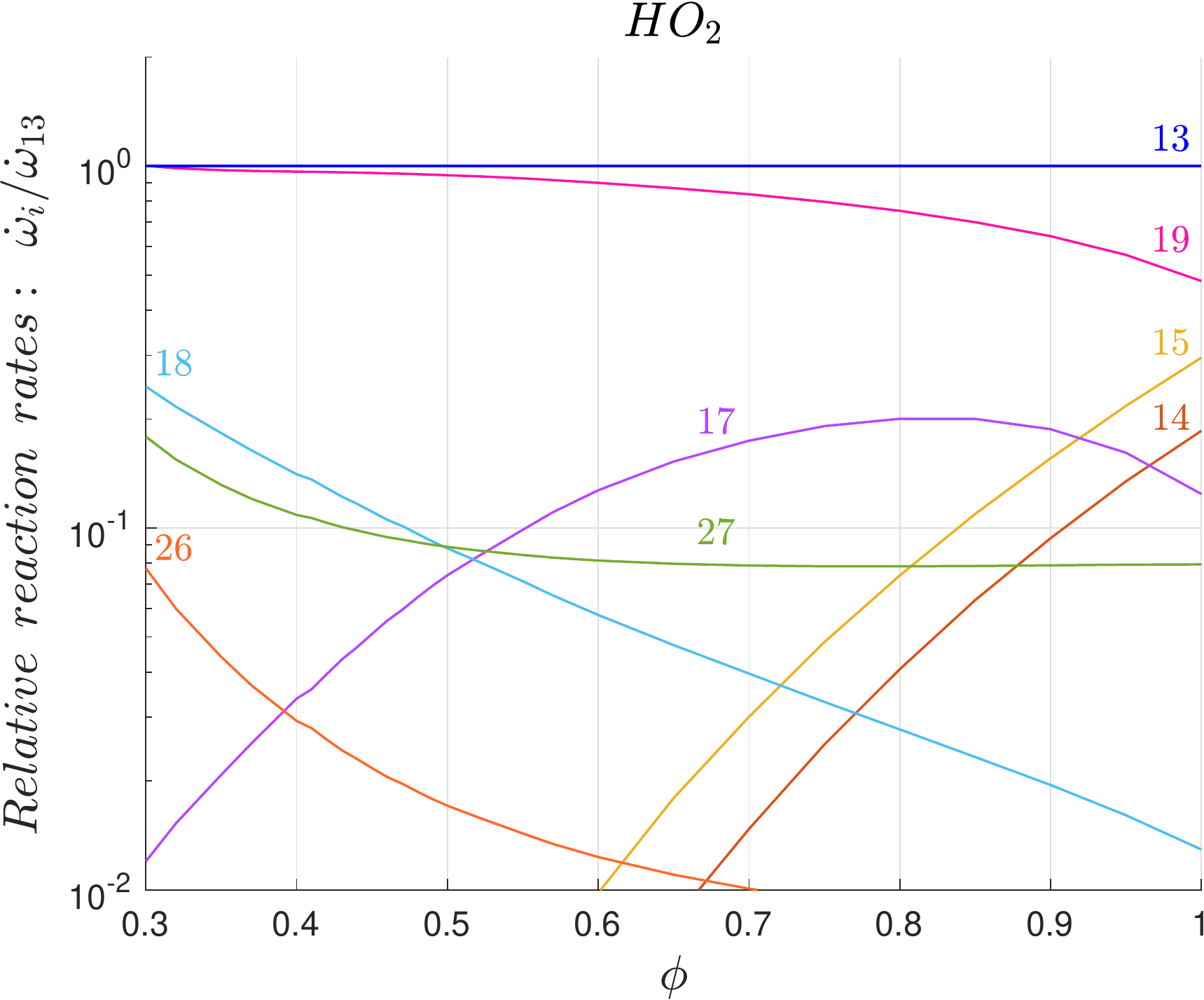} 
	\caption{Scaled rates of the relevant reactions of the \ce{HO2} kinetics (the numbers refer to the mechanism in Appendix D
	). The net consumption rate of \ce{HO2} is of the order of $\dot{\mathcal{N}}_{\ce{HO2}}/\dot{\omega}_{13} \sim 10^{-4}$, meaning that \ce{HO2} is well in steady state. The rates are evaluated at $x = 20$ mm, as in Figure \ref{H2Kinetics} where the choice is explained in detail.}
	\label{HO2SteadyState}
	\end{center}
\end{figure}

This form of the rates is based on the steady state of \ce{HO2} (see Figure \ref{HO2SteadyState}), which has the form
\begin{gather}
	\label{HO2SteadyStateAppendix}
	\C{\ce{HO2}}^{e} = 
	\frac{ \dot{\omega}^e_{13,f} + \sum_q {\dot{\omega}^e_{q,r}} }
		{ \dot{\omega}'^e_{13,r} + \sum_q {\dot{\omega}'^e_{q,f}} }, 
\end{gather}
obtained from the balance, $\dot{\omega}^e_{13} = \sum_{q\ne13} {\dot{\omega}^e_{q}}$, between the \ce{HO2} net production and consumption rates. The sums include all the \ce{HO2} reactions $q$ except reaction 13, and  the primed rates are evaluated without \C{HO2}.


To derive \eqref{HO2Ratesm} for $m \ne 13$ first write $\dot{\omega}_{m} \approx \dot{\omega}'^{e}_{m, f} \C{HO2}^{e} - \dot{\omega}^{e}_{m,r}$, 
%
which with \eqref{HO2SteadyStateAppendix}  gives
%
\begin{gather}
	\label{BasicRateHO2}
	\dot{\omega}_{m} / (\dot{\omega}'^{e}_{13,r} \mathcal{D} )\approx
		\dot{\omega}'^{e}_{m,f} \dot{\omega}^{e}_{13,f} 
			- \dot{\omega}^{e}_{m, r} \dot{\omega}'^{e}_{13,r} 
		+ 
		\sum_q { ( \dot{\omega}'^{e}_{m,f} \dot{\omega}^{e}_{q,r}  - \dot{\omega}^{e}_{m, r}\dot{\omega}'^{e}_{q,f}} ) ,
\end{gather}
%
with $\mathcal{D}$ defined below in \eqref{HO2RatesExpressions}. 

Using the equilibrium concentrations \eqref{PartialEquilibriumSolvedOnlyH2} yields $\dot{\omega}^{e}_{m, r} \dot{\omega}'^{e}_{13,r}/(\dot{\omega}'^{e}_{m,f} \dot{\omega}^{e}_{13,f}) \approx z^ 2$, so the numerator of the first term is $\dot{\omega}'^{e}_{m,f} \dot{\omega}^{e}_{13,f}(1 - z^2)$; the rest of the terms vanish since $\dot{\omega}^{e}_{p,f}/\dot{\omega}^{e}_{p,r} \propto z^{-1/2}$ for any $p \neq 13$, leading to \eqref{HO2Ratesm} with $\tilde{\Gamma}_m(z)$ given, for $m \ne 13$, by
%
%
\begin{subequations}
	\label{HO2RatesExpressions}
	%
	\begin{gather}
		\label{DinHO2Rates}
		\tilde{\Gamma}_m = \frac{\gamma_m }{\mathcal{SD}} \le 1, \\
		\label{gammas}
		\gamma_{m} = \dot{\omega}'^{e}_{m, f}/\dot{\omega}'^{e}_{13, r} \approx \gamma^b_{m} /z^{\alpha_m/2},  \\
		\mathcal{SD} = 1 +  \sum_m{\gamma_{m}},
	\end{gather}
and with $\alpha_{19} = 1$, $\alpha_{17} = 2$,  $\alpha_{14, 15} = 3$ for the reactions represented in Figure \ref{BigGammasWith1415App}, which also depicts these approximations.

The \ce{HO2} steady state gives $\dot{\omega}_{13}$ in turn with the same form \eqref{HO2Ratesm} with $\tilde{\Gamma}_{13} =  \sum_{m\ne13}{\tilde{\Gamma}_{m}} = (\mathcal{SD} -1)/\mathcal{SD}$.
\end{subequations}

\section{Shuffle reaction rates.} \label{ShuffleRatesApp}

{\it{Evaluation of $\tilde{\Gamma}_{4}$.}}
The three middle equations of \eqref{MainSystemFinal} give
\begin{subequations}
	\label{ShuffleRatesSolved}
	\begin{gather*}
		\tilde{\omega}_{1} \approx
			\frac{  
				\tilde{\mathscr{S}}^e_{\ce{OH}} \acute{X}^{e}_{\ce{H}} 
				- (\tilde{\mathscr{S}}^e_{\ce{O}} + \tilde{\mathscr{S}}^e_{\ce{H}} ) \acute{X}^{e}_{\ce{OH}} 
				+  (\tilde{\mathscr{S}}^e_{\ce{OH}} + \tilde{\mathscr{S}}^e_{\ce{H}} ) \acute{X}^{e}_{\ce{O}} 
			}
			{ \acute{X}^{e}_{\ce{OH}} + 2(\acute{X}^{e}_{\ce{O}} + 1) + 3\acute{X}^{e}_{\ce{H}} },	\\
		\tilde{\omega}_{4} \approx
			\frac{  \tilde{\mathscr{S}}^e_{\ce{OH}} + 2\tilde{\mathscr{S}}^e_{\ce{O}} + 3\tilde{\mathscr{S}}^e_{\ce{H}} }
			{ \acute{X}^{e}_{\ce{OH}} + 2(\acute{X}^{e}_{\ce{O}} + 1) + 3\acute{X}^{e}_{\ce{H}} },	 \\
		\tilde{\omega}_{5} \approx
			\frac{  
				\tilde{\mathscr{S}}^e_{\ce{OH}} \acute{X}^{e}_{\ce{H}} 
				- (\tilde{\mathscr{S}}^e_{\ce{O}} + \tilde{\mathscr{S}}^e_{\ce{H}} ) \acute{X}^{e}_{\ce{OH}} 
				+  (\tilde{\mathscr{S}}^e_{\ce{OH}} + \tilde{\mathscr{S}}^e_{\ce{H}} ) \acute{X}^{e}_{\ce{O}} 
			}
			{ \acute{X}^{e}_{\ce{OH}} + 2(\acute{X}^{e}_{\ce{O}} + 1) + 3\acute{X}^{e}_{\ce{H}} },
	\end{gather*}
\end{subequations}
with the rates $\tilde{\mathscr{S}}^e_{i}$ proportional to the \ce{HO2} reaction rates. For instance, with the reduced mechanism of Table \ref{KineticMechanism}, $\tilde{\mathscr{S}}^e_{\ce{O}} = -\tilde{\omega}_{17}$, $\tilde{\mathscr{S}}^e_{\ce{OH}} = \tilde{\omega}_{19} -\tilde{\omega}_{17}$ and $\tilde{\mathscr{S}}^e_{\ce{H}} = -\tilde{\omega}_{13}$, which, with \eqref{HO2RatesExpressions}, give $\tilde{\omega}_{4} \approx \tilde{\Gamma}_4(z) (1 - z^2)$, where
\begin{gather}
	\label{Gamma4}
	\tilde{\Gamma}_4(z) \approx
		\frac{  \tilde{\Gamma}_{19} - \tilde{\Gamma}_{17} - 2\tilde{\Gamma}_{17} + 3(\tilde{\Gamma}_{17} + \tilde{\Gamma}_{19}) }
		{ \acute{X}^{e}_{\ce{OH}} + 2(\acute{X}^{e}_{\ce{O}} + 1) + 3\acute{X}^{e}_{\ce{H}} }.
\end{gather}



%
\begin{figure}[h!]
	\begin{center}
	\includegraphics[width=0.44\textwidth]{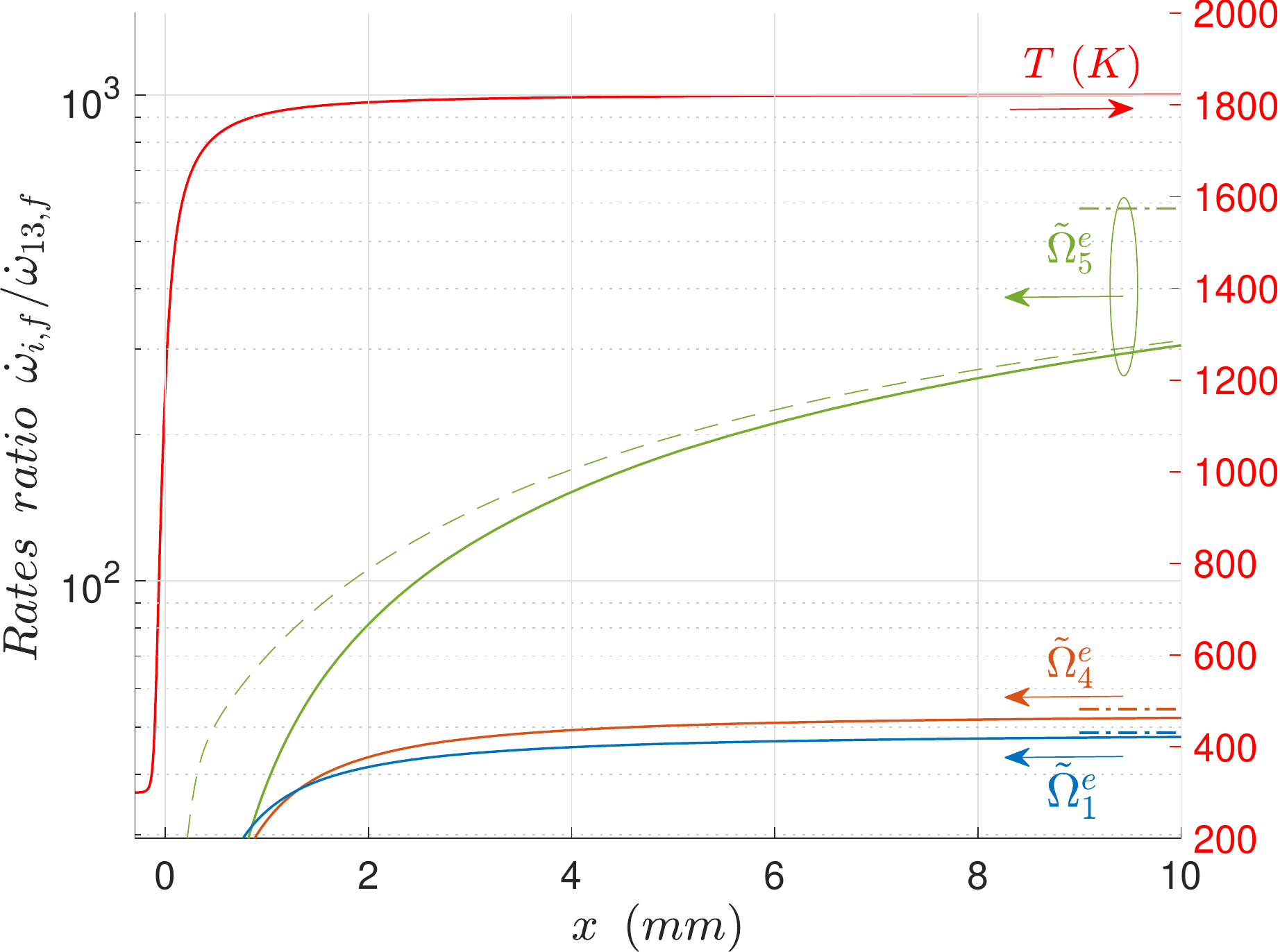} \hspace{1cm}
	\includegraphics[width=0.44\textwidth]{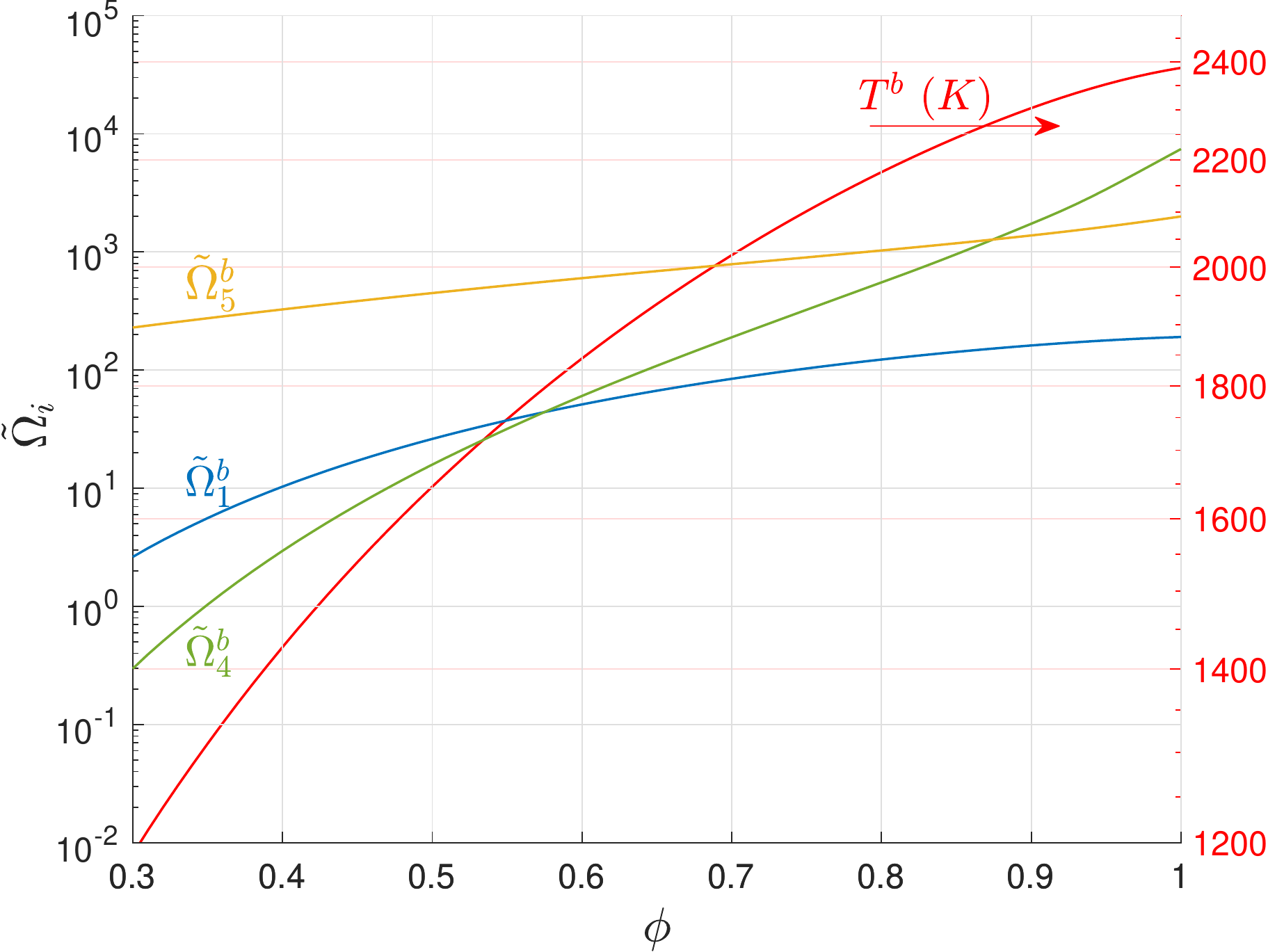}
	\caption{On the left, the distribution of the reaction rates ratios $\tilde{\Omega}^e_i$ for a flame with $\phi = 0.6$, showing that they soon reach their asymptotic values $\tilde{\Omega}^b_1$ and $\tilde{\Omega}^b_4 $, plotted in the right graph as functions of $\phi$. $\tilde{\Omega}^e_{5}$ has instead the form $\tilde{\Omega}^e_5 = \tilde{\Omega}^b_5\sqrt{z}$ (thin green dash line), with the asymptotic value $\tilde{\Omega}^b_5$ also represented in the right plot. Note that $\tilde{\Omega}^e_5 \gg 1$, so reaction 5 is in partial equilibrium for all $\phi < 1$. Reactions 1 and 4 however become out of partial equilibrium at the low-$\phi$ limit (see Figure \ref{PartialEquilibrium}).}
	\label{Omegas}
	\end{center}
\end{figure}

{\it{Deviations from zero net rates.}} The non-vanishing rates of the shuffle reaction are obtained in the approximation beyond partial equilibrium. The approach is exemplified first with reaction 4, with a rate $\dot{\omega}_{4} = k_{4, f} \C{H2}\C{OH} - k_{4, r} \C{H2O}\C{H}$, 
which gives at equilibrium $\dot{\omega}^{e}_{4,f} = k_{4, f}^b \C{H2}\C{OH}^e = (k_{4, r} \C{H2O})|^b\C{H}^e = \dot{\omega}^{e}_{4,r}$. Taking $\dot{\omega}^{e}_{4,f} = \dot{\omega}^{e}_{4,r}$ as common factor leads to
\begin{gather}
	\dot{\omega}_4/\dot{\omega}^e_{4,f} = \frac{k_{4, f}}{k_{4, f}^b}\psi_{\ce{OH}} - \frac{k_{4, r}\C{H2O}}{(k_{4, r}\C{H2O})|^b}\psi_{\ce{H}} ,
\end{gather}
%
in terms of the reduced concentrations $\psi_{s} = \C{s}/\C{s}^e$. 

Finally, approximating $\C{H2O}/\C{H2O}^b$ and the ratios $k_{4, (f,r)}/k_{4, (f,r)}^b$ by unity and dividing by $\dot{\omega}^{e}_{13,f} = (k_{13,f}\C{O_2})|^{b}\C{H}^{e}$ gives
\begin{subequations}
	\label{OmegaTildesApp}
	\begin{gather}
		\tilde{\omega}_{4} = \tilde{\Omega}^e_4 \left(\psi_{\ce{OH}} - \psi_{\ce{H}} \right),
	\end{gather}
	and similarly for reactions 1 and 5 
	\begin{gather}
		\tilde{\omega}_{-1} = \tilde{\Omega}^e_{-1} \left( \psi_{\ce{O}} \psi_{\ce{OH}} - \psi_{\ce{H}} \right), \\
		\tilde{\omega}_{5} = \tilde{\Omega}^e_5 \left( \psi_{\ce{OH}}^2 - \psi_{\ce{O}} \right) ,
	\end{gather}
\end{subequations}
with $\tilde{\Omega}^e_{i} = \dot{\omega}^{e}_{i,f}/\dot{\omega}^{e}_{13,f}$ functions of $z$ plotted in Figure \ref{Omegas}.

Using \eqref{PartialEquilibriumSolvedOnlyH2} shows 
that $\tilde{\Omega}^e_1$ and $\tilde{\Omega}^e_4$ are, in the first approximation, constants; 
whereas, 
$\tilde{\Omega}^e_{5} \approx \tilde{\Omega}^b_{5} \sqrt{z}$, as Figure \ref{Omegas} confirms in both cases. 

\section{Glarborg et al.'s mechanism.}\label{FulH2O2Mechanism}

The following two tables list the subsets of the Glarborg et al.'s \ce{H2}-\ce{O2} kinetic mechanism\cite{GLARBORG201831} used in this work. The reactions involving N have been removed. This introduces only negligible, non-essential numerical differences compared with the exact calculations with the full mechanism. The numbers used in the text are as shown here.

\begin{table}[h!]
\begin{tabular}{l c r}
	\fontsize{10pt}{14pt}\selectfont
	\begin{tabular}{l r @{} c @{} l }
			Label & & Reaction &  \\
			\midrule 
			 \multicolumn{2}{l}{ \hspace{.1cm} \fontsize{10pt}{13pt}\selectfont \ce{H2}-\ce{O2} subset}  &   & \\
			\cmidrule(l){1-2}
			1 &  \ce{H} + \ce{O2} &$\lra$ & \ce{O} + \ce{OH}  \\
			2, 3 &  \ce{O} +  \ce{H2} &$\lra$ & \ce{OH} + \ce{H}  \\
			4 &  \ce{OH} +  \ce{H2} &$\lra$ & \ce{H2O} + \ce{H}  \\
			5, 6 &  \ce{OH} + \ce{OH}  &$\lra $ & \ce{H2O} + \ce{O}  \\
			6 &  \ce{H2O} + \ce{M} &$\lra$ & \ce{H} + \ce{OH} + \ce{M}' \\
			7 &  \ce{H2} + \ce{M} &$\lra$ & \ce{H} + \ce{H} + \ce{M} \\
			8 &  \ce{H2} + \ce{Ar} &$\lra$ & \ce{H} + \ce{H} + \ce{Ar} \\
			9 &  \ce{H} + \ce{O} + \ce{M} &$\lra$ & \ce{OH} + \ce{M} \\
			10 &  \ce{O} + \ce{O} + \ce{M} &$\lra$ & \ce{O2} + \ce{M}' \\
			11 &  \ce{H2O} + \ce{M} &$\lra$ & \ce{H} + \ce{OH} + \ce{M}' \\
			12 &  \ce{H2O} + \ce{H2O} &$\lra$ & \ce{H} + \ce{OH} + \ce{H2O} \\
			\midrule 
	 \end{tabular}
&
\hspace{1cm}
&
	\begin{tabular}{l r @{} c @{} l }
			Label & & Reaction &  \\
			\midrule 
			\multicolumn{2}{l}{ \hspace{.1cm} {\fontsize{10pt}{14pt}\selectfont \ce{HO_2}-\ce{H2O2} subset}}  &   & \\
			\cmidrule(l){1-2}
			13 &  \ce{H} +  \ce{O2} +  \ce{M} &$\lra$ &  \ce{HO2} + \ce{M}  \\
			14 &  \ce{HO2} +  \ce{H} &$\lra$ & \ce{H2} + \ce{O2}  \\
			15 &  \ce{HO2} +  \ce{H} &$\lra$ & \ce{OH} + \ce{OH} \\
			16 & \ce{HO2} + \ce{H} &$\lra$ & \ce{H2O} + \ce{O} \\
			17 &  \ce{HO2} +  \ce{O} &$\lra$ &  \ce{OH} + \ce{O2}  \\
			18 &  \ce{HO2} +  \ce{OH} &$\lra$ & \ce{H2O} + \ce{O2} \\
			19 &  \ce{HO2} +  \ce{OH} &$\lra$ & \ce{H2O} + \ce{O2}  \\
			20 &  \ce{HO2} +  \ce{HO2} &$\lra$ & \ce{H2O2} + \ce{O2} \\
			21 &  \ce{HO2} +  \ce{HO2} &$\lra$ & \ce{H2O2} + \ce{O2} \\
			22 &  \ce{H2O2} +  \ce{M} &$\lra$ & \ce{OH} + \ce{OH} \\
			23 &  \ce{H2O2} +  \ce{H} &$\lra$ & \ce{H2O} + \ce{OH} \\
			24 &  \ce{H2O2} +  \ce{H} &$\lra$ & \ce{HO2} + \ce{H2} \\
			25 &  \ce{H2O2} +  \ce{O} &$\lra$ & \ce{HO2} + \ce{OH} \\
			26 &  \ce{H2O2} +  \ce{OH} &$\lra$ & \ce{HO2} + \ce{H2O} \\
			27 &  \ce{H2O2} +  \ce{OH} &$\lra$ & \ce{HO2} + \ce{H2O} \\
			\midrule \\
	 \end{tabular}
\end{tabular}
	\caption{Glarborg et al.'s \ce{H2}-\ce{O2} kinetic mechanism\cite{GLARBORG201831}.} 
	\label{FulH2O2MechanismTable}	
\end{table}

\bibliography{RecombinationRegionH2FlamesGrana}

\begin{thebibliography}{10}
\expandafter\ifx\csname url\endcsname\relax
  \def\url#1{\texttt{#1}}\fi
\expandafter\ifx\csname urlprefix\endcsname\relax\def\urlprefix{URL }\fi
\expandafter\ifx\csname href\endcsname\relax
  \def\href#1#2{#2} \def\path#1{#1}\fi

\bibitem{ZeldovichBook}
Y.~B. Zel{'}dovich, G.~I. Barenblatt, V.~B. Librovich, G.~M. Makhviladze, The
  mathematical theory of combustion and explosions, Consultants Bureau, 1985.

\bibitem{WilliamsBook}
F.~A. Williams, Combustion theory, The Benjamin/Cummings Publishing Company,
  Inc., 1985.

\bibitem{Linhan74}
A.~Li{\~n}{\'a}n, The asymptotic analysis of counterflow diffusion flames for
  large activation energies, Acta Astronaut. 1 (1974) 1007--1039.

\bibitem{SivashiskyHydrodynamics}
G.~I. Sivashinsky, On a distorted flame front as a hydrodynamic discontinuity,
  Acta Astronaut. 3 (1976) 889--918.

\bibitem{SeshadriReview}
K.~Seshadri, Multistep asymptotic analyses of flame structures, Proc. Combust.
  Inst. 26 (1996) 831--846.

\bibitem{WilliamsHydrogenSanchez}
A.~L. S\'anchez, F.~A. Williams, Recent advances in understanding of
  flammability characteristics of hydrogen, Prog. Energy Combust. Sci 41 (2014)
  1--55.

\bibitem{DaniHydrogen}
D.~Fern{\'a}ndez-Galisteo, A.~L. S{\'a}nchez, A.~Li{\~n}{\'a}n, F.~A. Williams,
  The hydrogen--air burning rate near the lean flammability limit, Combust.
  Theor. Model. 13~(4) (2009) 741--761.

\bibitem{DaniHydrogenI}
D.~Fern{\'a}ndez-Galisteo, A.~L. S{\'a}nchez, A.~Li{\~n}{\'a}n, F.~A. Williams,
  One-step reduced kinetics for lean hydrogen-air deflagration, Combust. Flame
  156 (2009) 985--996.

\bibitem{seshadriMethane}
K.~Seshadri, N.~Peters, The inner structure of methane-air flames, Combust.
  Flame 81~(2) (1990) 96 -- 118.

\bibitem{SeshadriMethaneII}
M.~Bui-Pham, K.~Seshadri, F.~A. Williams, The asymptotic structure of premixed
  methane-air flames with slow {CO} oxidation., Combust. Flame 89 (1992)
  343--362.

\bibitem{LinanLeanMethane}
A.~L. S{\'a}nchez, A.~L{\'e}pinette, M.~Bollig, A.~Li{\~n}{\'a}n, B.~Lazaro,
  The reduced kinetic description of lean premixed combustion, Combust. Flame
  123 (2000) 436--464.

\bibitem{GRANAOTERO2019115750}
J.~Gra{\~n}a-Otero, S.~Mahmoudi, Excited {OH} kinetics and distribution in
  \ce{H_2} premixed flames, Fuel 255 (2019) 115750.

\bibitem{cantera}
D.~G. Goodwin, R.~L. Speth, H.~K. Moffat, B.~W. Weber, Cantera: An
  object-oriented software toolkit for chemical kinetics, thermodynamics, and
  transport processes, \url{https://cantera.org/}, version 2.4.0 (2018).

\bibitem{GLARBORG201831}
P.~Glarborg, J.~A. Miller, B.~Ruscic, S.~J. Klippenstein, Modeling nitrogen
  chemistry in combustion, Progress in Energy and Combustion Science 67 (2018)
  31 -- 68.

\bibitem{SanDiego}
F.~A. Williams, K.~Seshadri, R.~J. Cattolica, {S}an {D}iego {M}echanism,
  \url{http://combustion.ucsd.edu} (2014).

\bibitem{KONNOV201914}
A.~A. Konnov, Yet another kinetic mechanism for hydrogen combustion, Combust.
  Flame 203 (2019) 14 -- 22.

\bibitem{LinanRecombination}
A.~Li{\~n}{\'a}n, A theoretical analysis of premixed flame propagation with an
  isothermal chain reaction, AFOSR Contract No. E00AR68-0031 (1971).

\bibitem{Zeldovich1961TwoSteps}
Y.~B. Zeldovich, Kinetika i Kataliz 2 (1961) 305.

\bibitem{LinanWilliams}
A.~Li{\~n}{\'a}n, F.~A. Williams, Fundamental Aspects of Combustion, Oxford
  University Press, 1993.

\bibitem{HITRAN}
\href{https://www.cfa.harvard.edu/hitran/}{The {HITRAN} {D}atabase} [online].
\newblock Accessed: 07-05-2018.

\bibitem{COELHO2018105}
F.~R. Coelho, F.~H. Fran{\c c}a, {WSGG} correlations based on {HITEMP2010} for
  gas mixtures of \ce{H2O} and \ce{CO2} in high total pressure conditions,
  International Journal of Heat and Mass Transfer 127 (2018) 105 -- 114.

\bibitem{JOULIN1979139}
G.~Joulin, P.~Clavin, Linear stability analysis of nonadiabatic flames:
  Diffusional-thermal model, Combustion and Flame 35 (1979) 139 -- 153.

\end{thebibliography}
\bibliographystyle{elsarticle-num}

\end{document}